\documentclass{aa}  

\usepackage{graphicx}
\usepackage{txfonts}
\usepackage{natbib}
\usepackage{graphicx}
\usepackage{txfonts}
\usepackage[normalem]{ulem}
\usepackage{xcolor} 
\usepackage[dvipsnames]{xcolor} 

\usepackage{placeins} 

\usepackage{caption}
\usepackage{placeins} 

\definecolor{links}{rgb}{0.11, 0.67, 0.84}
\usepackage[colorlinks=true,allcolors=links]{hyperref}

\newcommand{\Msun}[1]{$\mathrm{M}_\odot$}
\newcommand{\Mstar}[1]{$\mathrm{M}_\star$}
\newcommand{\ntracer}[1]{$n_\mathrm{tracer}$}
\newcommand{\zerr}[1]{$z_\mathrm{err}$}
\newcommand{\zbest}[1]{$z_\mathrm{best}$}

\begin{document}

   \title{Unveiling the small-scale web around galaxies with miniJPAS and DESI}

    \titlerunning{Unveiling the small-scale web around galaxies}

   \subtitle{}

   \author{Daniela Gal\'arraga-Espinosa\inst{\ref{MPA}, \ref{IPMU}}\thanks{danigaes@mpa-garching.mpg.de, daniela.galarraga@ipmu.jp}
          \and
          Guinevere Kauffmann\inst{\ref{MPA}}
          \and
          Silvia Bonoli\inst{\ref{DIPC}, \ref{Ikerbasque}}
          \and
          Luisa Lucie-Smith\inst{\ref{Hamburg}}
          \and
          Rosa M. González Delgado\inst{\ref{IAA}}
          \and 
          Elmo Tempel\inst{\ref{Tartu}, \ref{Estonia}}
          \and 
          Raul Abramo\inst{\ref{InstFisicaSaoPaulo}}
          \and
          Siddharta Gurung-López\inst{\ref{OAValencia}, \ref{UValencia}}
          \and 
          Valerio Marra\inst{\ref{UEspiritoSanto}, \ref{INAF}, \ref{IFPU}}
          \and 
          Jailson Alcaniz\inst{\ref{ObsRioJaneiro}}
          \and
          Narciso Benitez
          \and
          Saulo Carneiro \inst{\ref{ObsRioJaneiro}}
          \and
          Javier Cenarro \inst{\ref{CEFCA}, \ref{CEFCAIAA}}
          \and
          David Cristóbal-Hornillos \inst{\ref{CEFCA}}
          \and
          Renato Dupke \inst{\ref{ObsRioJaneiro}}
          \and
          Alessandro Ederoclite \inst{\ref{CEFCA}, \ref{CEFCAIAA}}
          \and
          Antonio Hernán-Caballero \inst{\ref{CEFCA}, \ref{CEFCAIAA}}
          \and
          Carlos Hernández-Monteagudo \inst{\ref{IAC}, \ref{ULL}}
          \and 
          Carlos López-Sanjuan \inst{\ref{CEFCA}, \ref{CEFCAIAA}}
          \and 
          Antonio Marín-Franch \inst{\ref{CEFCA}, \ref{CEFCAIAA}}
          \and
          Claudia Mendes~de~Oliveira \inst{\ref{USaoPaulo}}
          \and
          Mariano Moles \inst{\ref{CEFCA}}
          \and
          Laerte Sodré~Jr \inst{\ref{USaoPaulo}}
          \and
          Keith Taylor \inst{\ref{Instruments4}}
          \and
          Jesús Varela \inst{\ref{CEFCA}}
          \and 
          Hector Vázquez~Ramió \inst{\ref{CEFCA}, \ref{CEFCAIAA}} 
          }

   \institute{Max-Planck Institute for Astrophysics, Karl-Schwarzschild-Str.~1, D-85741 Garching, Germany\label{MPA}
    \and
    Kavli IPMU (WPI), UTIAS, The University of Tokyo, Kashiwa, Chiba 277-8583, Japan\label{IPMU}
    \and
    Donostia International Physics Center (DIPC), Paseo Manuel de Lardizabal 4, 20018 Donostia-San Sebastian, Spain \label{DIPC}
    \and 
    IKERBASQUE, Basque Foundation for Science, E-48013, Bilbao, Spain \label{Ikerbasque}
    \and
    Hamburger Sternwarte, Universit{\"a}t Hamburg, Gojenbergsweg 112, D-21029 Hamburg, Germany \label{Hamburg}
    \and
    Instituto de Astrofísica de Andalucía (IAA-CSIC), P.O.\~Box 3004, 18080 Granada, Spain\label{IAA}
    \and
    Tartu Observatory, University of Tartu, Observatooriumi 1, 61602 Tõravere, Estonia\label{Tartu}
    \and
    Estonian Academy of Sciences, Kohtu 6, 10130 Tallinn, Estonia \label{Estonia} 
    \and
    Departamento de Física Matemática, Instituto de Física, Universidade de São Paulo, Rua do Matão, 1371, CEP 05508-090, São Paulo, Brazil \label{InstFisicaSaoPaulo}
    \and
    Observatori Astron\`omic de la Universitat de Val\`encia, Ed. Instituts d’Investigaci\'o, Parc Cient\'ific. C/ Catedr\'atico Jos\'e Beltr\'an, n2, 46980 Paterna, Valencia, Spain \label{OAValencia}
    \and
    Departament d’Astronomia i Astrof\'isica, Universitat de Val\`encia, 46100 Burjassot, Spain \label{UValencia}
    \and
    Departamento de Física, Universidade Federal do Espírito Santo, 29075-910, Vitória, ES, Brazil \label{UEspiritoSanto}
    \and
    INAF -- Osservatorio Astronomico di Trieste, via Tiepolo 11, 34131 Trieste, Italy \label{INAF}
    \and
    IFPU -- Institute for Fundamental Physics of the Universe, via Beirut 2, 34151, Trieste, Italy \label{IFPU}
    \and 
    Observatório Nacional, Rua General José Cristino, 77, São Cristóvão, 20921-400, Rio de Janeiro, RJ, Brazil \label{ObsRioJaneiro}
    \and 
    Centro de Estudios de Física del Cosmos de Aragón (CEFCA), Plaza San Juan 1, E--44001, Teruel, Spain \label{CEFCA}
    \and 
    Unidad Asociada CEFCA-IAA, CEFCA, Unidad Asociada al CSIC por el IAA, Plaza San Juan 1, 44001 Teruel, Spain\label{CEFCAIAA}
    \and
    Instituto de Astrofísica de Canarias (IAC), C. Vía Láctea, 38205 La Laguna, Santa Cruz de Tenerife, Spain \label{IAC}
    \and
    Universidad de La Laguna, Avda Francisco Sánchez, E-38206, San Cristóbal de La Laguna, Tenerife, Spain \label{ULL}
    \and
    Universidade de São Paulo, Instituto de Astronomia, Geofísica e Ciências Atmosféricas, Rua do Matão, 1226, 05508-090, São Paulo, SP, Brazil \label{USaoPaulo}
    \and 
    Instruments4, 4121 Pembury Place, La Canada Flintridge, CA 91011, U.S.A \label{Instruments4}
    }

   \date{Received XXX; accepted YYY}

  \abstract{We present the first statistical observational study detecting filaments in the immediate surroundings of galaxies, i.e. the local web of galaxies. Simulations predict that cold gas, the fuel for star formation, is channeled through filamentary structures into galaxies. Yet, direct observational evidence for this process has been limited by the challenge of mapping the cosmic web at small scales. Using miniJPAS spectro-photometric data combined with spectroscopic DESI redshifts when available, we construct a high-density observational galaxy sample spanning $0.2 < z < 0.8$. Local filaments are detected within a 3 Mpc physical radius of each galaxy with stellar mass \Mstar~$>10^{10}$~\Msun~ using all nearby galaxies as tracers, combined with a probabilistic adaptation of the DisPerSE algorithm designed to overcome limitations due to photometric redshift uncertainties. Our methodology is tested and validated using mock catalogues built with random forest models applied to a simulated lightcone. Besides recovering the expected increase in galaxy connectivity (defined as the number of filaments attached to a galaxy) with stellar mass, we show that our connectivity measurements agree with 3D reference estimates from the mock galaxies. Thanks to these filament reconstructions, we explore the relation between small-scale connectivity and galaxy star formation rate, finding a mild positive trend which needs to be confirmed by follow up studies with larger sample sizes. We propose galaxy connectivity to local filaments as a powerful and physically motivated metric of environment, offering new insights into the role of cosmic structure in galaxy evolution.}

   \keywords{galaxies: evolution -- galaxies: star formation -- galaxies: fundamental parameters -- large-scale structure of Universe -- methods: data analysis -- methods: statistical}

   \maketitle
%

\section{Introduction}

Understanding how galaxies stop forming stars, and what drives this process, remains one of the key open questions in galaxy evolution. Observationally, galaxies in the late Universe ($z \ll 2$) tend to be either actively star-forming or already quenched, with only a small fraction caught in between \citep{Strateva2001, Baldry2004, Schawinski2014}. Several mechanisms have been proposed to explain this star formation quenching, ranging from internal processes like AGN and supernova feedback \citep[e.g.][]{Silk1998, DiMatteo2005Nature, Singh2020, Piotrowska2022} to external environmental effects such as gas stripping or strangulation in dense environments \citep[e.g.][]{Gunn1972, Moore1996Nature, Kauffmann2004, Peng2010_QuenchingGalaxies}.

While there is ongoing debate about whether internal or external mechanisms dominate star formation quenching, and it is likely that both contribute with varying importance, a crucial and often overlooked piece of the puzzle is the origin of the fuel itself. Galaxies need access to cold gas in order to keep forming stars \citep[][]{Prescott2015_Lya_largescales, Zabl2019_Lya}. Cosmological simulations predict that this gas is funnelled along filaments of the multiscale cosmic web \citep[e.g.][]{Keres2005, Dekel2009_coldstreams, Nelson2013_movingmesh, FaucherGiguere2011_streams_and_outflows, Mandelker2018}. The accessibility of these filamentary streams is therefore expected to play a key role in regulating galaxy growth \citep[e.g.][]{Borzyszkowski2017_ZOMG1, RomanoDiaz2017_ZOMG2, Garaldi2018_ZOMG3, Lyu2025ApjL_gasaccretion_coplanar}. In this context, a natural pathway towards quenching could be the loss or disconnection from these cold gas supply channels, for example, when galaxies become dynamically disconnected from their local filamentary network, as proposed in the cosmic web detachment scenario \citep{AragonCalvo2019_CWdisconnection}, or when the shear flow in the galaxy's environment limits the accretion of matter onto the halo \citep{Borzyszkowski2017_ZOMG1}. 

Nevertheless, most observational studies exploring the galaxy-filament connection have focused on the more easily observable large-scale cosmic filaments, i.e. the backbones of the cosmic web at the largest-scales, connecting galaxy clusters and at the intersections of cosmic walls and voids \citep{White1987, Bond1996}. These include works based on data from GAMA, SDSS, Euclid, and other wide-field galaxy surveys \citep[e.g.][]{Alpaslan2016_GAMAgalaxies, Malavasi2017, Laigle2018, Malavasi2020_sdss, Welker2020_sami, Laigle2025_Euclid}. Results remain mixed, with some studies suggesting that large-scale filaments enhance star formation \citep{Fadda2008, Darvish2014, Kleiner2017, Vulcani2019}, while others find evidence for quenching \citep{Martinez2016, Chen2017_fil_gal, Kraljic2018, Bonjean2020filaments, Castignani2022}. Most recently, however, \citet{Okane2024} showed that, once over-density effects are carefully controlled for, large-scale filaments have no significant impact on star formation \citep[see also][]{Navdha2025}. This result is unsurprising given the gas properties of large-scale filaments at low redshift, where most of these studies have been carried out. Indeed, studies in hydro-dynamical simulations have shown that these structures are principally filled with warm, diffuse gas \citep{GalarragaEspinosa2021, Tuominen2021} at $z=0$, a gas phase that has been detected in X-rays experiments \citep[up to redshift of about 0.6,][]{Tanimura2020_Rosat, Tanimura2022_eRosita, Zhang2024erosita_fils}, thus unlikely to fuel star formation.\footnote{We note that the picture at higher redshifts is more complex than the one at $z \sim 0$. For example, at $z \sim 2$, studies in simulations have found a non-negligible influence of the large-scale filaments having an effect on some galaxy properties \citep{Song2021, GalarragaEspinosa2023}.} Still, if a galaxy evolved inside e.g.~ a thick filament, one would expect some measurable imprint on its evolution compared to a galaxy that grew up a region with weaker tidal forces, like a void or a thin filament, as discussed in \cite{Borzyszkowski2017_ZOMG1}. However,
such an influence is unlikely to affect the instantaneous star formation rate, and would rather manifest as an integrated effect that built up over the lifetime of the galaxy.

To make progress in understanding quenching, galaxy evolution studies should instead focus on filamentary structures at smaller-scales, i.e. those filaments arising from the local density field around galaxies \citep{Pichon2010,AragonCalvo2010, GalarragaEspinosa2023}, in the physical region that can actually influence galaxy evolution, with matter density properties that make them much more likely to carry cold gas \citep{Ramsoy2021}. To our knowledge, no observational studies have yet systematically explored the connection between galaxies and these structures, largely due to the high spatial resolution, tracer density, and redshift accuracy required to reconstruct the local web in the environments of galaxies. For example, previous studies of galaxy connectivity, such as the pioneering work of \cite{Kraljic2020}, have used filaments traced by SDSS galaxies which, by selection, are relatively bright and sparse galaxies. As a result, the filaments they identify correspond to large-scale cosmic structures \citep[cosmic filaments of Mpc widths, ][]{Wang2024fil_radius} rather than to the local, small-scale filaments we aim at probing in this paper \citep[of widths of about tens of kiloparsecs,][]{Ramsoy2021}.\\

In this work, we take a major step forward by presenting the first systematic detection of local filamentary structures around galaxies across the stellar mass range, \Mstar~ $= 10^{10} - 10^{11.74}$ \Msun~, and over the redshift range $0.2 < z < 0.8$. We combine the spectro-photometric data from miniJPAS with the spectroscopic precision of DESI redshifts (when available) to build a dense galaxy sample suitable for tracing the small-scale, local web. Using a probabilistic filament detection method, we reconstruct the filamentary structures within a 3 Mpc proper (pMpc) radius around each galaxy, using lower mass and other galaxies of the catalogue as tracers. By working in proper (or physical) rather than comoving coordinates, we ensure that the physical size of the probed environment remains consistent across redshift. This is a crucial point, as we want to detect structures that are physically connected to the galaxies, and because processes like gas accretion, interactions, or stripping act on physical scales.

Within this framework, we use galaxy connectivity ($K$), defined as the number of filaments connected to a galaxy, as our principal environmental metric. Traditional descriptors like over-density or group membership are useful but miss key information about the geometry and anisotropy of the field. Connectivity, in contrast, offers a complementary approach as it captures the channels through which gas, satellites \citep[][]{Welker2018, Madhani2025}, and angular momentum are delivered \cite[e.g.][]{Pichon2011, Danovich2012_streams, Kraljic2020,Gouin2021}.
The structure of this paper is as follows. In Sect.~\ref{Sect:jpas+desi_catalogue}, we present the miniJPAS+DESI galaxy catalogue. Sect.~\ref{Sect:mocks} describes the construction of the mock catalogue used for validation. The methods employed in this work are introduced in Sect.~\ref{Sect:Method}. We present our main results in Sect.~\ref{Sect:Results} and discuss their implications in Sect.~\ref{Sect:discussion}. Finally, Sect.~\ref{Sect:Conclusions} summarises our findings and conclusions.

\section{\label{Sect:jpas+desi_catalogue}The miniJPAS+DESI galaxy catalogue}

\begin{figure} 
    \centering
    \includegraphics[width=0.95\linewidth]{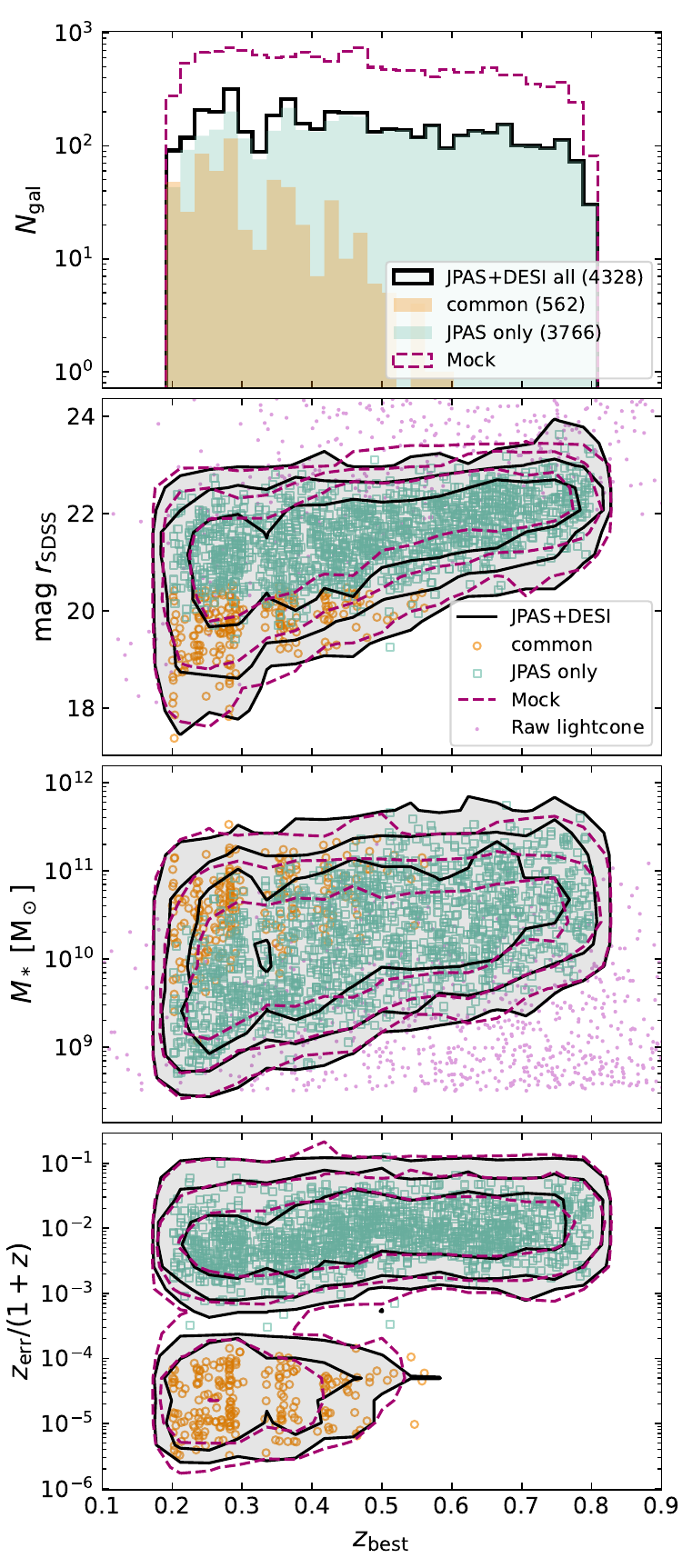}
    \caption{Main properties of the JPAS+DESI catalogue (4328 galaxies) as a function of redshift. From top to bottom, we show the number of galaxies, the $r_\mathrm{SDSS}$ band magnitudes, stellar masses, and redshift precision distributions. 
    The contours enclose 68, 95, and 99.7\% of the data points, respectively, from the inner to the outer lines. Black solid lines show the properties of the observed JPAS+DESI catalogue (Sect.~\ref{SubSubSect:final_catalogue}) while purple dashed ones correspond to the mock catalogue introduced in Sect.~\ref{Sect:mocks}. For illustration, pink dots show a random subsample of the raw lightcone galaxies from which the mocks were derived.
    Orange and blue colours correspond respectively to galaxies from the matched (common) catalogue and from the J-PAS only dataset. The scattered points illustrate a random subsample of 1500 galaxies from the total JPAS+DESI catalogue.}
    \label{Fig:catalogue_properties}
\end{figure}

In this section, we present the different steps and selections done to build the observational galaxy catalogue used in this work, namely the JPAS+DESI catalogue.

\subsection{\label{SubSubSect:miniJPAS}miniJPAS galaxies}

We use public data of the miniJPAS survey \citep{Bonoli2021_miniJPAS}, a precursor of the J-PAS survey, covering $\sim 1 \, \mathrm{deg}^2$ of the sky using the J-PAS unique set of 56 photometric filters. Among those, 54 are overlapping narrow bands with a full-width half-maximum of $\sim 145 \, \AA$, and two are broader band filters in the blue and red extremes of the optical spectrum \cite[see Fig.~2 of][]{Bonoli2021_miniJPAS}. This narrowband filter system is specifically designed to deliver accurate photometric redshift measurements. Indeed, the miniJPAS survey provides an effective low resolution spectrum (equivalent to a resolution of R $\sim 60$) for every object  detected, thus overcoming typical spectroscopic survey limitations such as target selection biases and fiber collision issues. This allows for the detection of fainter objects and a higher sampling density compared to typical spectroscopic surveys.

The galaxy selection cuts done in this work are stated in the following. After discarding flagged objects using both \texttt{flags} and \texttt{mask\_flags} fields \citep{Bonoli2021_miniJPAS}, we focus on the objects with \texttt{class\_star} $\leq 0.1$ and \texttt{odds} $\geq 0.6$. While the former selects objects classified as galaxies, the latter ensures that those galaxies have good photo-spectra and thus reliable photo-$z$ values. Indeed, the \texttt{odds} parameter \citep{Benitez2000} is defined as the integral of the redshift PDF within the integration range $2 \times 0.03 \, (1 + z)$ centred at the mode of the PDF \citep[for further details see][]{Bonoli2021_miniJPAS, HernanCaballero2021_JPAS}. Values of \texttt{odds} range from 0 to 1, where higher values indicate photo-$z$ estimates with greater confidence. The lower limit of \texttt{odds} $\geq 0.6$ used in this work was determined through a systematic analysis of catalogues with different \texttt{odds} cuts, ranging from $0.3$ to $0.9$. Our results (not shown) confirm that galaxies with \texttt{odds} $< 0.6$ exhibit catastrophic redshift errors \citep[in agreement with][]{HernanCaballero2021_JPAS}. On the other hand, imposing a higher \texttt{odds} threshold becomes too restrictive, significantly reducing the number of galaxies, particularly at $z > 0.4$. Thus, we adopt \texttt{odds} $= 0.6$ as the optimal balance between photo-$z$ accuracy and sample statistics for our analysis. This \texttt{odds} cut selects $36.5\%$ of the galaxies. 
To effectively probe the local environments of \Mstar~>$10^{10}$ \Msun~ galaxies, we require a field of view large enough to encompass regions with a 3 pMpc radius centred on each of those target galaxies. This spatial constraint sets a lower redshift limit of 0.2 given the limited footprint of miniJPAS. We also set an upper redshift limit of 0.8. This was instead determined after examination of the evolution of the galaxy number density of the catalogue, which sharply drops below $10^{-3}$ comoving $\mathrm{Mpc}^{-3}$ beyond $z = 0.8$, making it increasingly challenging to detect small-scale structures at these redshifts.

Throughout this paper, we use the \texttt{MAG\_AUTO} miniJPAS magnitudes, which are estimated using a Kron-like elliptical aperture \citep{Bonoli2021_miniJPAS}. We use Kron-like elliptical magnitudes because they provide a more consistent estimate of total galaxy flux by adapting to the object's light profile and shape, reducing biases from fixed-aperture measurements \cite{Kron1980}.
Redshifts are computed using the Lephare code \citep{Arnouts2011_lephare}, following \cite{HernanCaballero2021_JPAS}. 
Galaxy properties, namely the stellar masses and star-formation rates (SFR), are taken from \cite{GonzalezDelgado2021}. These were derived by fitting the galaxies spectral energy distributions (SEDs), traced by the 56 J-PAS narrowband filters, using the Bayesian parametric code BaySeAGal. For details about this SED-fitting code, we refer the reader to \cite{GonzalezDelgado2021}.

Finally, to mitigate border effects in the filament detection, we filled the small masked regions (mainly caused by bright stars) of the miniJPAS footprint with randomly distributed points until the density of tracers of the catalogue was matched. In addition, we added a 3 pMpc buffer region along the survey edges. These artificial points serve only to stabilise the filament finding procedure by preventing survey gaps from being misidentified as voids and survey borders from being falsely interpreted as density ridges.

\subsection{\label{SubSubSect:matchDESI}Matching with DESI}

We improve the accuracy of the photometric redshifts of some galaxies
in the miniJPAS sample by matching the latter with data from the DESI spectroscopic survey. 
Precisely, we use redshifts from the DESI Bright Galaxy Survey \citep[hereafter DESI BGS,][]{Hahn2022_DESIBGS, Hahn2023_DESI_bgs} derived from the DESI early data release \citep{Adame2024_DESI_EDR}. 
This DESI BGS catalogue\footnote{\url{https://data.desi.lbl.gov/doc/releases/edr/vac/provabgs/}} is magnitude-limited to $\sim 20.2$ in the $r$-band, and represents about 1\% of the final DESI Main Survey data. It includes two distinct galaxy samples (bright and faint), both used in this work, spanning a redshift range from 0 to 0.6. 
After removing galaxies with bad measurements (\texttt{MAG\_R} and \texttt{provabgs\_logMstar\_bf} = -999.0) and imposing positive values for the signal-to-noise metric for observed spectra (\texttt{TSNR2\_BGS} > 0.0), we match the positions of the DESI galaxies to the miniJPAS ones using a matching radius of $10^{-3.5}$ degrees (1.14 arcseconds). This value was selected based on the study of 2D sky separations between DESI and miniJPAS galaxies, whose bimodal distribution showed a net separation at this value.

In the following, \zbest~ refers to the best available redshift for each galaxy, taken from DESI spectroscopy when available, or otherwise from the high accuracy miniJPAS photometry. We point out that, even for the galaxies in common with DESI, we use the stellar masses and SFRs estimates from miniJPAS in order to maintain a consistent dataset across all redshifts.

\subsection{\label{SubSubSect:final_catalogue}Final JPAS+DESI catalogue}

We have combined the miniJPAS and DESI datasets to build the JPAS+DESI catalogue containing 4328 galaxies in the redshift range $[0.2, 0.8]$. Among these, only the brighter and lower redshift galaxies (562) are common between miniJPAS and DESI, while the vast majority (3766) come from the miniJPAS dataset, which extends to fainter magnitudes and higher redshifts. 
This is clearly seen in the different panels of Fig.~\ref{Fig:catalogue_properties}. From top to bottom, we show the redshift distribution and the distribution of $r$-band magnitudes, stellar masses, and redshift precision as a function of $z_\mathrm{best}$ (the best galaxy redshift). Redshift precision is here defined as the relative redshift error: $z_\mathrm{err} / (1 + z)$. 

In the bottom-most panel, we appreciate the significant improvement of redshift precision thanks to the DESI contribution. Indeed, replacing the photo-$z$ value by spec-$z$ information for the common galaxies (orange points) resulted in a decrease of several orders of magnitude in the redshift errors. For these galaxies, the contour, centred at $z_\mathrm{err} / (1 + z) \sim 10^{-5}$, is completely disjoint from the typical values of the J-PAS only galaxies (teal points). 
We estimated that a line-of-sight (l.o.s.) uncertainty of 3 pMpc or less corresponds to a redshift precision of roughly $z_\mathrm{err} / (1 + z) \lesssim 6 \times 10^{-4}$ in this redshift range. While DESI comfortably meets this requirement, J-PAS alone does not achieve this precision. Nevertheless, DESI provides only sparse spatial sampling, which is why J-PAS data are essential for this work, as it enables denser coverage required to probe the local environments of galaxies. All the details of the method for filament reconstruction using this dataset will be provided in Sect.~\ref{Sect:Method}.

\subsection{\label{SubSubSect:target_galaxies}Definition of target galaxies}

We focus our filament detection efforts on galaxies with stellar masses above $10^{10}$~\Msun~, hereafter referred to as target galaxies. 
This lower mass limit was chosen based on completeness considerations. The analyses of \cite{GonzalezDelgado2021} and \cite{DiazGarcia2024} have shown that the $50\%$ mass completeness of miniJPAS (at $r_\mathrm{SDSS} \sim 22.5$ AB) is \Mstar~ = $10^{9.0-9.2}$ \Msun~ at $z = 0.2$, and $10^{10.2-10,6}$ \Msun~ at z = 0.6, depending on galaxy colour. Our adopted threshold of $10^{10}$~\Msun~ thus ensures that the low redshift targets are safely above the completeness limit, while the higher redshift ones remain close to it without sacrificing too much statistics.

To reliably probe the $3$ pMpc environments radially around these targets, we limit our sample to galaxies located at least 3 pMpc from the edges of the miniJPAS footprint. We also ensure that the targets lie far from the redshift boundaries of the catalogue by requiring their redshifts to leave a 15 pMpc buffer within the range $z = [0.2, 0.8]$. This value is justified later in Sect.~\ref{Subsect:LocalEnvSelection}. These additional selection criteria are designed to avoid border effects that could bias the filament reconstruction. After these cuts, our sample contains 2272 target galaxies.

In Sect.~\ref{SubSubSect:mock_validation}, we evaluate the number density of tracer galaxies surrounding the target galaxies. Based on the filament reconstruction performance tests, we restrict our analysis to environments with a tracer density above 0.0117~$\mathrm{pMpc}^{-3}$, corresponding to at least 10 tracers in the cylindrical volume of $3^2\pi \times (2\times15) \, \, \mathrm{pMpc}^3$. The final sample includes 1547 target galaxies that satisfy this criterion. We have checked that these selected galaxies possess essentially the same stellar mass, magnitude, and redshift distributions as those with lower tracer densities, indicating that this selection does not introduce any noticeable selection bias.

\section{\label{Sect:mocks}Mock observations}

In this study, we pay particular attention to the calibration of the methods and to the assessment of the detected filaments. To this end, it is essential to compare our results to mock observations that closely resemble the observed JPAS+DESI catalogue. 

We build mock observations using the simulated lightcone\footnote{\url{https://galformod.mpa-garching.mpg.de/public/LGalaxies/downloads.php}} from \cite{Henriques2015_Lgal}. This lightcone, based on the L-galaxies model applied to the Millennium simulation, is suitable for mocking our JPAS+DESI catalogue because it matches observed galaxy properties such as masses and colours, with a mass resolution limit down to $10^7$ \Msun~, which is low enough for modelling even the lowest masses of our JPAS+DESI catalogue. 
In the following, we present the different steps involved in building the mocks. We first mimic the selection function of the observed catalogue using a random forest algorithm (Sect.~\ref{SubSubSect:RFC}), we then match the number densities (Sect.~\ref{SubSubSect:Mock_density_match}), and model the observed photometric and spectroscopic redshift errors (Sect.~\ref{SubSubSect:Mock_zerr}). Finally, we present our mock validation in Sect.~\ref{SubSubSect:mock_validation}.

\subsection{\label{SubSubSect:RFC} Random forest classifier}

In order to mimic the JPAS+DESI selection function in our mock catalogues and to account for intrinsic correlations between observed properties, we train a random forest classifier on the J-PAS data to distinguish between galaxies with \texttt{odds} $<0.6$ (\texttt{class0}) and \texttt{odds} $\geq 0.6$ (\texttt{class1}).
The input features for the model were selected based on the galaxy properties that correlate the most with the \texttt{odds} parameter. These are redshifts, apparent magnitudes ($r_\mathrm{SDSS}$, $i_\mathrm{SDSS}$, $g_\mathrm{SDSS}$ bands), and colours ($g$-$r$, $r$-$i$, and $g$-$i$). The training was performed on the J-PAS catalogue with a minimal odds cut of 0.3 to exclude galaxies with very bad measurements, which would not be present in the simulation. The dataset was split into $70\%$ training and $30\%$ testing. 
The inspection of the resulting Receiver Operating Characteristic (ROC) curve (not shown), which quantifies the performance of the classifier at different probability thresholds, showed that a probability threshold of $0.74$ ensures a contamination level below $10\%$ for \texttt{class1} galaxies. The resulting completeness (of 61\%) is less critical as the simulated galaxies are later randomly under-sampled to match the observed catalogue number density (see next).

\subsection{\label{SubSubSect:Mock_density_match} Application to the lightcone and density matching}

We applied the trained model to the \cite{Henriques2015_Lgal} lightcone and selected the simulated galaxies with a probability of 0.74 or above of belonging to \texttt{class1}. This outputs a simulated galaxy catalogue that closely reproduces the intrinsic correlations between galaxy properties of our \texttt{odds} $\geq 0.6$ JPAS+DESI catalogue, while maintaining less than $10\%$ contamination. 
We then matched the observed number densities by randomly sampling the simulated galaxies. This matching was done in redshift bins to capture the slight decrease of the number density with redshift of our JPAS+DESI catalogue.

The resulting mock redshifts, $r_\mathrm{SDSS}$ magnitude, and stellar mass distributions are presented by the purple dashed lines in the first three panels of Fig.~\ref{Fig:catalogue_properties}. 
We find a very good agreement between the observations and the mocks, as all observed ranges and trends are well recovered. Two-sample KS tests also reveal no significant differences between the observed and mock distributions. Notably, the stellar mass distribution is remarkably well recovered, despite the model not being explicitly trained on this feature. This is a reassuring sign of the reliability and robustness of our model.

\begin{figure} 
    \centering
    \includegraphics[width=1\linewidth]{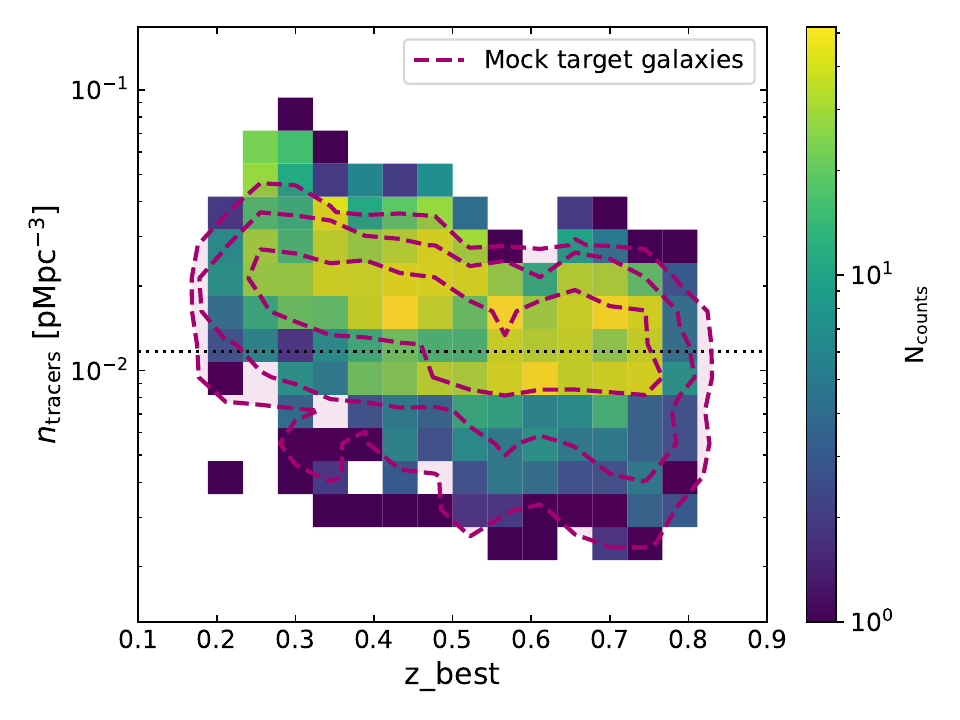}
    \caption{Distribution of the density of tracers \ntracer~ around observed and mock target galaxies as a function of redshift (Sect.~\ref{SubSubSect:mock_validation}). The colormap presents the results for the observed JPAS+DESI target galaxies, while purple dashed contours show the mock target galaxy distribution. Contours enclose 68, 95, and 99.7\% of the points. The black dotted horizontal line marks the 0.0117~$\mathrm{pMpc}^{-3}$ value (Sect.~\ref{SubSubSect:target_galaxies}). Target galaxies above this line have at least 10 tracers in their local environment.
    }
    \label{Fig:ntracers_dist_obs_and_mock}
\end{figure}

\subsection{\label{SubSubSect:Mock_zerr} Modelling redshift errors}

The last step in building our mock catalogue is modelling the redshift errors. After exploring the parameter space, we found that a simple $r_\mathrm{SDSS}$ magnitude cut at $\sim 20.45$ effectively separates the population of galaxies with JPAS-like photo-$z$ errors from the DESI-like one, having spec-$z$ errors. This magnitude threshold is consistent with the magnitude limit of the DESI catalogue used in this work. 
After dividing simulated galaxies based on this magnitude cut, we assign redshift errors to each one of them by sampling the $z_\mathrm{err}$ values from the corresponding observed distribution following a lognormal model (chosen after inspection of the observed distributions). This sampling was performed in redshift bins to account for the slight increase in photo-$z$ errors with redshift. The resulting mock redshift precision distribution is shown by the purple dashed lines in the last panel of Fig.~\ref{Fig:catalogue_properties}. Once again, we note a good agreement with the observations.

\subsection{\label{SubSubSect:mock_validation}Mock validation}

We test the ability of the mocks to replicate the specific environments of observed JPAS+DESI target galaxies. This is achieved by comparing the number density of tracers (\ntracer~) around mock target galaxies to that around observed target galaxies. Specifically, \ntracer~ is defined as the number of objects within a cylindrical volume of radius 3 pMpc and height $2\times 15$ pMpc, oriented parallel to the l.o.s. and centred on each target galaxy (mock or observed). 
Mock target galaxies are selected following the same criteria as in the observational case (see Sect.~\ref{SubSubSect:final_catalogue}).

The \ntracer~ distributions as a function of redshift are presented in Fig.~\ref{Fig:ntracers_dist_obs_and_mock}. We appreciate the good agreement between the mock and the observed results, confirming that our mocks do not only match the main properties of the JPAS+DESI catalogue, but also reproduce the environments of target galaxies, an essential requirement for this work. We note that the observed JPAS+DESI galaxies present a small number count excess towards the low redshift and high \ntracer~ values. This is due to the presence of over-dense structures (such as galaxy groups and clusters) identified in the miniJPAS field \citep{Maturi2023_jpasclusters}, which are absent in the lightcone due to its limited simulated sky coverage.


\section{\label{Sect:Method}Filament identification}

\begin{figure*} 
    \centering
    \includegraphics[width=1\linewidth]{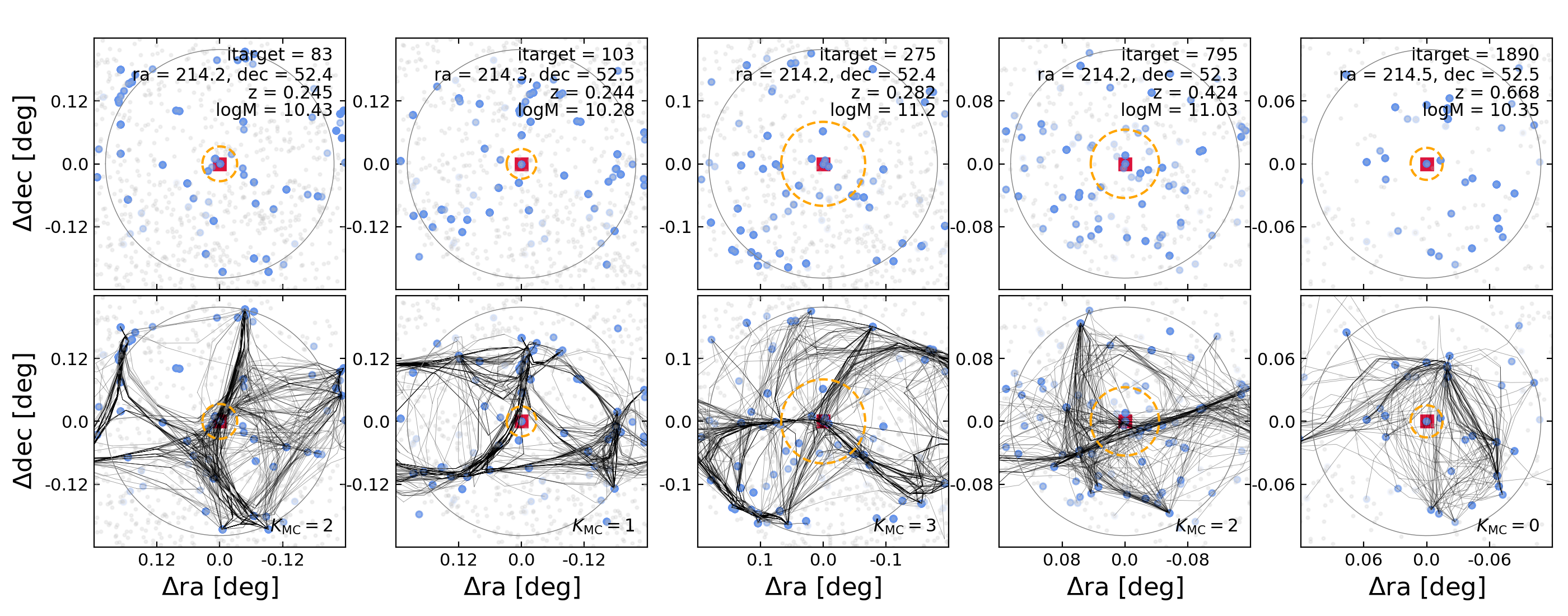}
    \caption{Examples of five different observed target galaxies (red central squares). Points in shades of blue correspond to their associated Monte Carlo (MC) galaxy distributions. The gray outer circles show the aperture of radius 3 pMpc at the redshift of the target. The gray dots in the background show the positions of all the JPAS+DESI galaxies (no selection made) for illustration purposes. In the bottom panels, the black thin lines, over-plotted on top of the same galaxy distributions, show the positions of the resulting filaments. Orange dashed circles mark the aperture of $20 \times R_1$ used for connectivity measurements (Sect.~\ref{SubSect:MeasuringK_method}). }
    \label{Fig:OBS_MCvisu}
\end{figure*}

Photometric redshift errors pose a major problem in the detection of filaments at all scales. This is because the subsequent uncertainty in the position of a galaxy along the l.o.s., hereafter the confusion length, defined as:
\begin{equation}\label{Eq:confusionlength}
    \Lambda_c = d_A(z - z_\mathrm{err}) - d_A(z + z_\mathrm{err}),
\end{equation}
where $d_A$ is the angular diameter distance, biases the identification of 3D structures such as filaments by blurring such structures. 
In this work, we adopt a novel approach to deal with this problem. This method, first introduced in \cite{Laigle2025_Euclid} and used in \cite{Gouin2025_Euclid} in the context of galaxy clusters studies, is based on probabilistic filament reconstructions through Monte Carlo (hereafter, MC) samplings of redshift probability distributions. Concretely, it allows galaxy redshift values to vary following a normal distribution\footnote{Given our galaxy selection based on a high value for the \texttt{odds} threshold (0.6), assuming normal redshift probability distributions is well justified. These galaxies with reliable measurements typically have relatively peaked and to normal-like $z$-PDFs.} centred on \zbest~ and with standard deviation \zerr~. Filaments are detected in each individual realisation of the catalogue, thus effectively propagating the redshift uncertainties into the reconstructed skeletons and subsequent statistics. In this context, filaments that are consistently detected across most realisations can be interpreted as `high confidence' filaments, while those detected in only a few realisations should be considered as noise or `low confidence' filaments. 
In the following, we present our adaptation of this method to the context of this work, where we focus on the local environments of galaxies.

\subsection{\label{Subsect:LocalEnvSelection}Selection of local environments around target galaxies}

We generate 100 random Monte Carlo realisations of the full JPAS+DESI catalogue. For each target galaxy in a given realisation, we select all the other galaxies within a cylindrical volume of radius 3 pMpc and height $2\times 15$ pMpc, centred on the target galaxy position and oriented along the redshift axis. 
The slice thickness of 30 pMpc was selected after analysing the confusion length (Eq.~\ref{Eq:confusionlength}) of our JPAS+DESI catalogue. We have estimated the mean confusion length to be $\Lambda_c \sim 20$ pMpc at $z=0.2$, and going up to $35$ pMpc at $z=0.8$, due to the larger photometric uncertainties at higher redshifts \citep{HernanCaballero2021_JPAS}. Therefore, the chosen slice thickness is designed to encompass the intrinsic redshift uncertainties of the catalogue.

This environment selection is repeated for all 100 realisations. Some examples of the resulting galaxy distributions around the targets are presented in the top panels of Fig.~\ref{Fig:OBS_MCvisu}, showing the 2D projections along the line of sight. Each Monte Carlo realisation is plotted with translucid blue points, so galaxies that are present in multiple realisations appear to be more opaque. The gray outer circles denote the 3 pMpc radius at the redshift of the target galaxy. Galaxies at slightly higher redshifts may appear outside the circles because of projection effects in sky coordinates.\footnote{Indeed, since the angular selection radius is fixed at the redshift of the target galaxy, in sky coordinates galaxies at high $z$ will appear outside the selection radius, although they lie physically within the 3D cylinder defined above.}

\subsection{\label{Subsect:disperse}DisPerSE application}

\begin{figure*}
    \centering
    \includegraphics[width=0.5\linewidth]{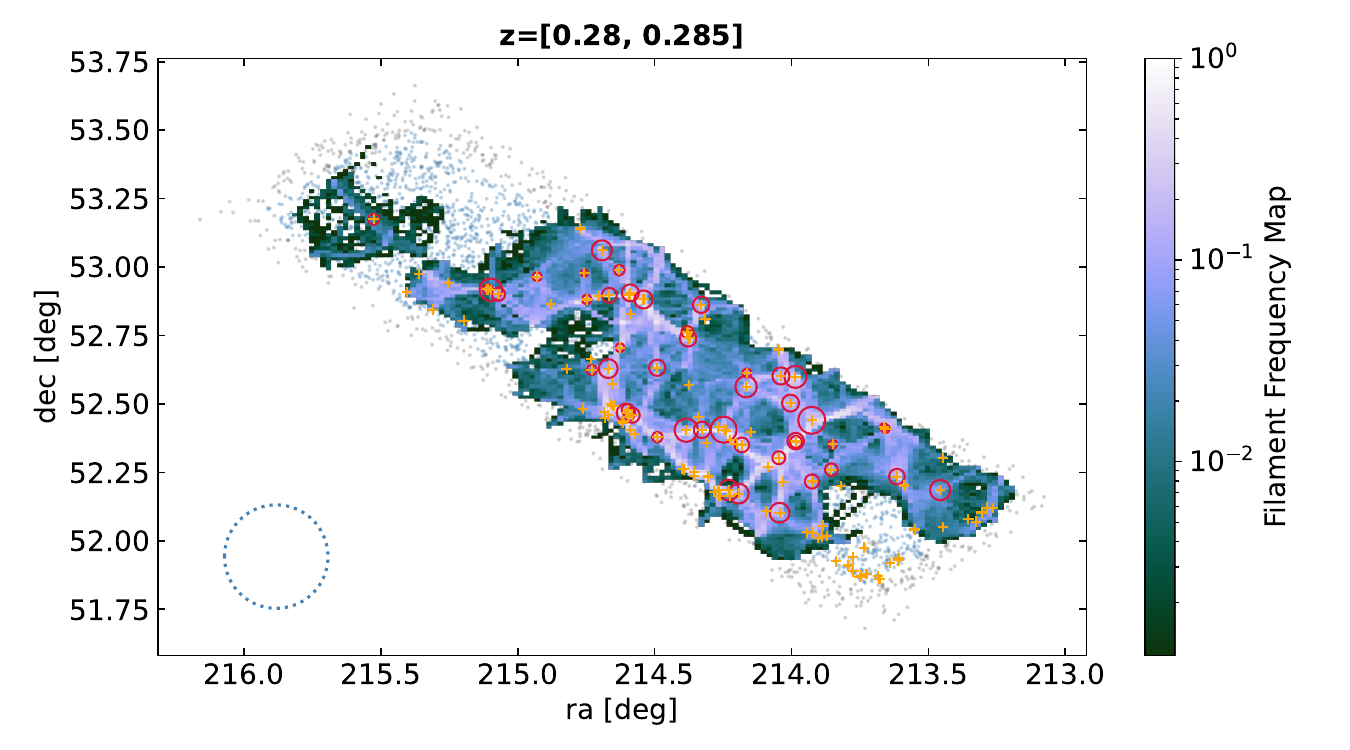}\includegraphics[width=0.5\linewidth]{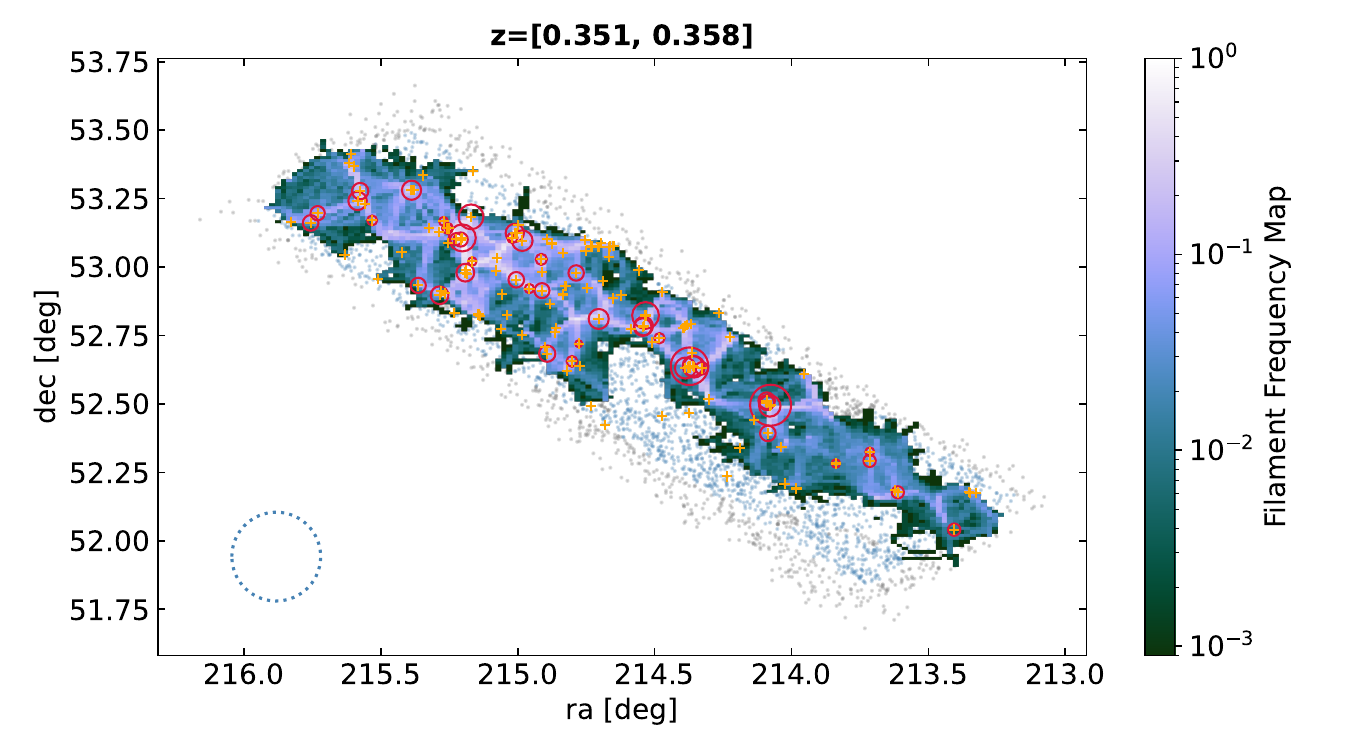}
    \includegraphics[width=0.5\linewidth]{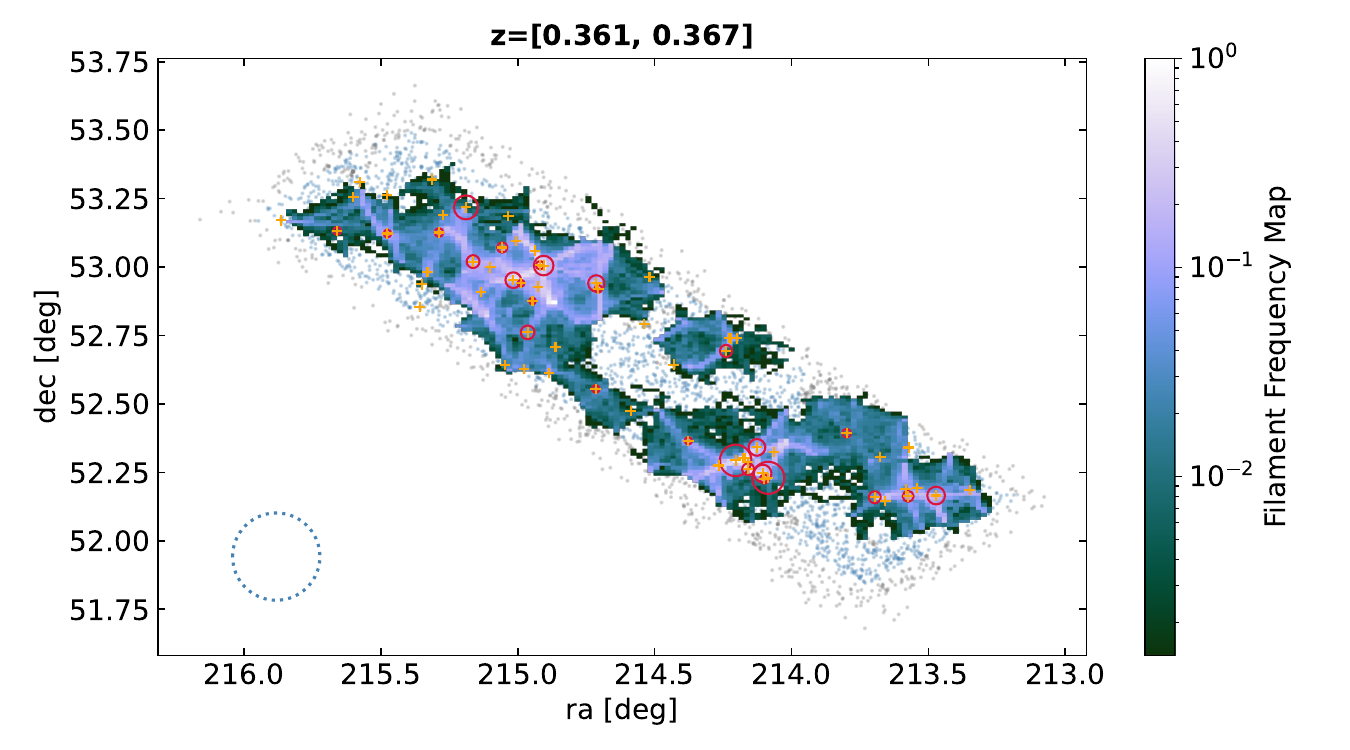}\includegraphics[width=0.5\linewidth]{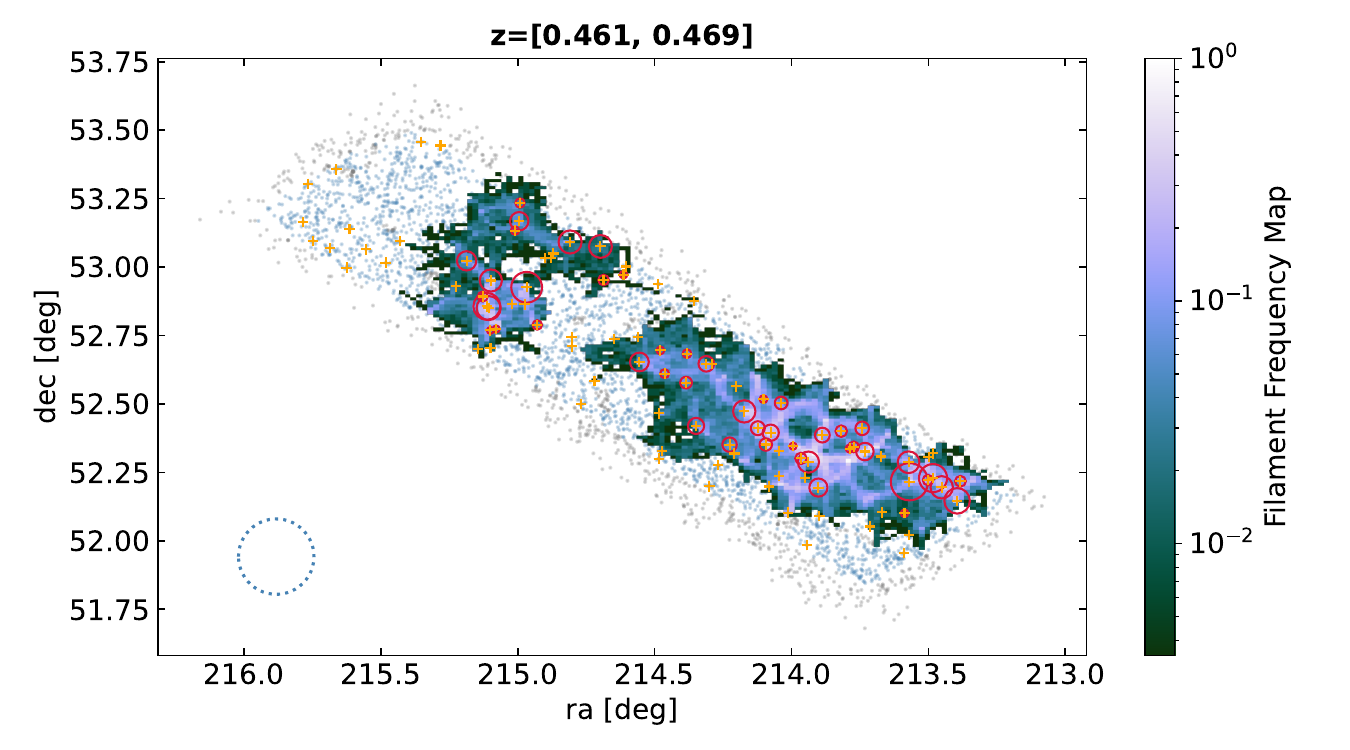}
    \caption{Filament frequency maps measured in our JPAS+DESI galaxy catalogue. Each panel corresponds to the projection of our data in a different redshift slice, each of equivalent thickness 12 pMpc (arbitrary choice, just for visualisation). Red circles show the target galaxies with sizes proportional to their stellar masses (Sect.~\ref{SubSubSect:target_galaxies}). Orange points correspond to tracer galaxies from a single realisation. Pixel colours indicate the number of filament detections (or hits) normalised by the maximum number of hits within the slice (Sect.~\ref{Subsect:filament_detection}). The dotted blue circle in the bottom left corner of the plots shows a scale of radius 3 pMpc measured at the mean redshift of the slice. Blue points in the background show the $\sim 1 , \mathrm{deg}^2$ footprint of the miniJPAS survey and gray points denote the artificial buffer region added at the survey edges, following Sect.~\ref{SubSubSect:miniJPAS}.
    }
    \label{Fig:filament_maps}
\end{figure*}

Filaments are detected using the DisPerSE code \citep{Disperse_paper1, Sousbie2011b}. This finder is designed to identify filaments by analysing the topology of the density field. In the present case, this is derived from the 2D projected spatial distribution of galaxies in the cylindrical volumes around the targets, using the Delaunay Tessellation Field Estimator \citep[DTFE,][]{SchaapWeygaert2000, WeygaertSchaap2009}.
DisPerSE identifies the peaks and saddle points of the field, and defines filaments as the sets of segments connecting these critical points, following the ridges of the field. 

It is fair to mention that we work under the implicit assumption that the spatial distribution of galaxies traces the underlying dark matter field, albeit with some unavoidable bias, as discussed in \cite{Zakharova2023} and \cite{Bahe2025_filaments} for larger-scale filaments. In the vicinity of haloes, analyses of high-resolution hydrodynamical simulations \citep[e.g. TNG50-1, ][]{Nelson2019_TNG50, Pillepich2019_TNG50} show that the galaxy distribution is not random but follows filamentary patterns aligned with the dark matter density field \cite[][Galárraga-Espinosa et al., in prep.]{GalarragaEspinosa2023}. These results indicate that galaxies broadly trace the structure of the dark matter potential wells, with filamentary patterns extending beyond the first-order spherical approximation of haloes. This supports our (observationally necessary) choice of using galaxies as tracers of the local filamentary network.

Following recent work \citep{Kuchner2020, Cornwell2023, Cornwell2024}, we mass-weight the vertices of the Delaunay tessellation before the filament identification step. In this technique, the density value of each vertex of the field, corresponding to a galaxy position, is weighted by the stellar mass of that galaxy. This weighting enhances the contrast of the underlying density field, giving more importance to massive galaxies that better trace the large-scale potential. As a result, the detection of physically meaningful topological nodes is improved, leading to more robust and reliable filamentary structures than those obtained with the standard, unweighted approach \citep{Kuchner2020}. 

DisPerSE's main free parameter is the persistence ratio (\texttt{-nsig}), defined as the density ratio between critical point pairs. This parameter must be carefully calibrated to ensure the robustness of filaments against unavoidable Poisson noise, characteristic of discrete distributions. Following the method introduced in Appendix C of \cite{GalarragaEspinosa2023}, we determine that the optimal DisPerSE persistence threshold for our dataset and method is $3\sigma$, as it effectively mitigates noise while preserving the signal. 
Finally, after filament detection, we apply a single smoothing step to the filament spines to reduce sharp angles. This adjustment is purely cosmetic, and we have verified that it does not affect our connectivity results.

\subsection{\label{Subsect:filament_detection}Probabilistic filament detection}

For each 1547 target galaxy, we apply DisPerSE to each of the 100 Monte Carlo realisations projected in 2D, representing a  computationally demanding task. 
Examples of the resulting observed filaments are shown in the bottom panels of Fig.~\ref{Fig:OBS_MCvisu}. The results illustrate the strength of our method, as filaments become clearly visible through the probabilistic approach. 
For a more global view, Fig.~\ref{Fig:filament_maps} presents the resulting filament reconstructions in some redshift slices. 
To produce this figure, we have binned all target galaxies into redshift slices of thickness equivalent to 12 pMpc (arbitrary choice for visualisation), projected their associated filaments into 2D grids, and computed filament frequency maps by calculating the number of filament detections (or hits) per pixel and normalising by the maximum number of hits within the slice.
These maps beautifully reveal the structure of the multiscale cosmic web, with a clear presence of nodes, filaments, and other emptier regions. This figure serves as a posteriori, qualitative check of our method, demonstrating that it successfully recovers a coherent web at fixed redshift even though filaments were detected locally and independently around each target galaxy. We note that this figure is provided for visualisation purposes only and that throughout this work we use the raw individual Monte Carlo reconstructions.

\subsection{\label{SubSect:MeasuringK_method}Measuring galaxy connectivities}

We define galaxy connectivity, $K$, as the number of filaments intersecting a circle of radius $20 \times R_1$ centred on each target galaxy, as illustrated by the orange dashed circles in Fig.~\ref{Fig:OBS_MCvisu}. The $R_1$ radius, first introduced in \cite{Trujillo2020_R1radius}, is a physically motivated measure of galaxy size, derived from the analysis of deep observations. It corresponds to the radial position of the stellar mass density isocontour at $1 \, M_{\odot} \, \mathrm{pc}^{-2}$, which is as a proxy for the outermost radius where gas has enough density to collapse and form stars \citep[see][and references therein]{Martinez-Lombilla2019, Trujillo2020_R1radius}.  
Here, we use the recent scaling relations from \cite{Arjona-Galvez2025} to estimate $R_1$ from galaxy stellar masses. These relations, validated in both observations and simulations, apply to galaxies across a broad stellar mass range, exhibit remarkably low scatter, are redshift independent, and robust against variations in baryonic models of state-of-the-art zoom-in simulations.

We choose to perform the galaxy connectivity measurements at a radius of $20 \times R_1$ because, based on our analysis of numerical simulations (not shown), this radius corresponds to $\sim 1/3$ of the host halo's virial radius, placing it within the circum\-galactic medium (CGM) regime. This choice ensures that connectivity measurements are adapted to the scales of each galaxy, preventing biases from mixing different spatial scales. For reference, the aperture radii used in this work range from $20 \times R_1 = 130$ proper kpc to 1.87 proper Mpc depending on galaxy mass, with a mean value of 481 kpc. 

\begin{figure}
    \centering
    \includegraphics[width=1\linewidth]{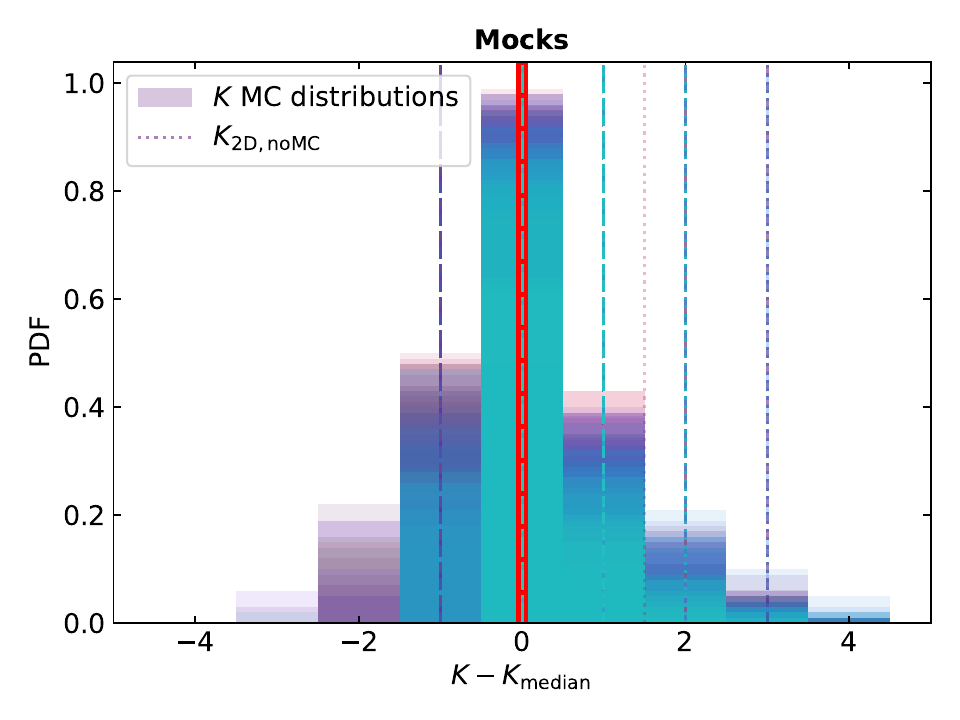}
    \includegraphics[width=1\linewidth]{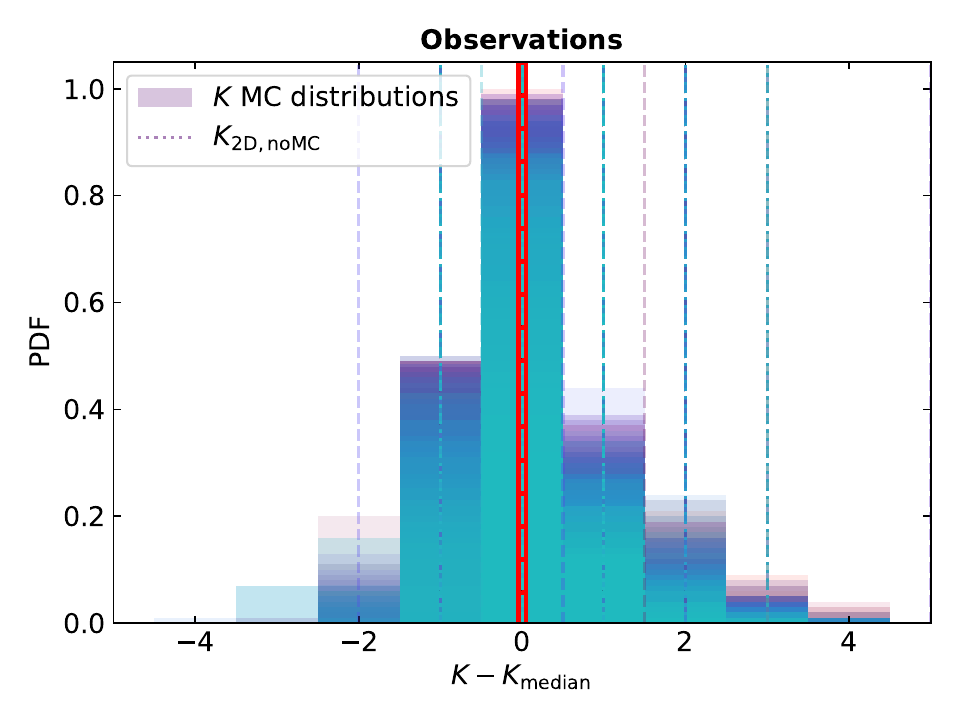}
    \caption{Connectivity distributions of the Monte Carlo realisations for 300 random galaxies (different colours in the mock and observational datasets, shown respectively in the top and bottom panels. Each distribution is rescaled by its median value and therefore centred at $x=0$ (thick vertical red line). The thin vertical lines with different line styles indicate the corresponding $K_\mathrm{2D,noMC}$ values, i.e. the results from a single measurement in the fiducial catalogue, without Monte Carlo sampling. These lines are often superimposed due to a quantisation effect: $K_\mathrm{2D,noMC}$ is an integer by definition; half-integer positions appear after rescaling when the median of the distribution is itself a half-integer.
    }
    \label{Fig:K_randomdist}
\end{figure}

For each target galaxy, we measure one connectivity value per Monte Carlo realisation. The final connectivity, hereafter $K_\mathrm{MC}$, is defined as the median across all realisations. For illustration, Fig.~\ref{Fig:K_randomdist} shows examples of the connectivity distributions for 300 random galaxies, each represented by a different colour, for the mock and observed catalogues (top and bottom panels, respectively). The distributions have been rescaled to be centred on their median value ($K_\mathrm{MC}$, by definition).
To demonstrate the strength of our approach, we indicate with vertical dotted lines the connectivity obtained from a single measurement on the 2D projected galaxy distribution of the fiducial catalogue (i.e. using the same galaxy selection around the target but without Monte Carlo resampling), hereafter $K_\mathrm{2D,noMC}$. These values have been rescaled by their corresponding $K_\mathrm{MC}$ to enable a direct comparison across galaxies. In many cases $K_\mathrm{2D,noMC}$ values are visibly offset from the peak of the corresponding distribution, illustrating their susceptibility to stochastic effects. The median based $K_\mathrm{MC}$ values are therefore more stable and statistically robust.

In Appendix~\ref{APP:apertureradius}, we present connectivity results obtained using different values for the aperture radii and also various fixed constant radii. From this analysis, we conclude that the results presented in this paper are qualitatively robust to the specific choice of aperture radius used to measure connectivity.

\begin{figure*}
    \centering
    \includegraphics[width=1\linewidth]{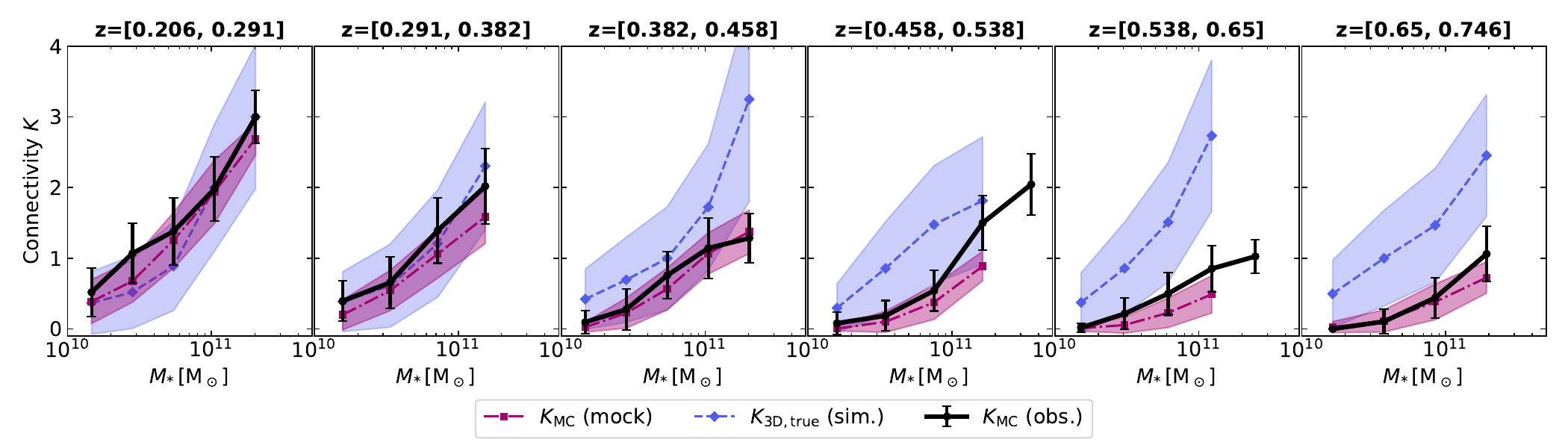}
    \caption{Galaxy connectivity as a function of stellar mass, shown for different redshift bins (columns). Redshift bin edges were chosen to contain approximately the same number of observed galaxies. The black, magenta, and blue curves correspond to the mean connectivity measured from the observed JPAS+DESI galaxies, the mock galaxies, and the ideal 3D connectivity measurements, respectively (see details in Sect.~\ref{SubSect:Kcomparison_mocks}). Averages based on fewer than five galaxies are not shown. The error bars correspond to the $1\sigma$ dispersion of the data points around the mean connectivity within a given mass bin. 
    }
    \label{Fig:Kcomparisons_1row}
\end{figure*}


\section{\label{Sect:Results}Results}

In this section, we present the results of our local filament reconstructions, focusing on connectivity as the primary metric to characterise the filamentary environments of target galaxies. In Sect.~\ref{SubSect:Kcomparison_mocks} we present the first ever observational galaxy connectivity measurements and interpret them by comparing with results from simulated catalogues. We further analyse the $K$ measurements in Sect.~\ref{SubSect:Kmeasurements}, and we finally show the results on its impact on galaxy star-formation rate in Sect.~\ref{SubSect:SFR}.

\subsection{\label{SubSect:Kcomparison_mocks}First connectivity results: comparison with reference measurements}

   \begin{figure}
   \centering
    \includegraphics[width=0.45\textwidth]{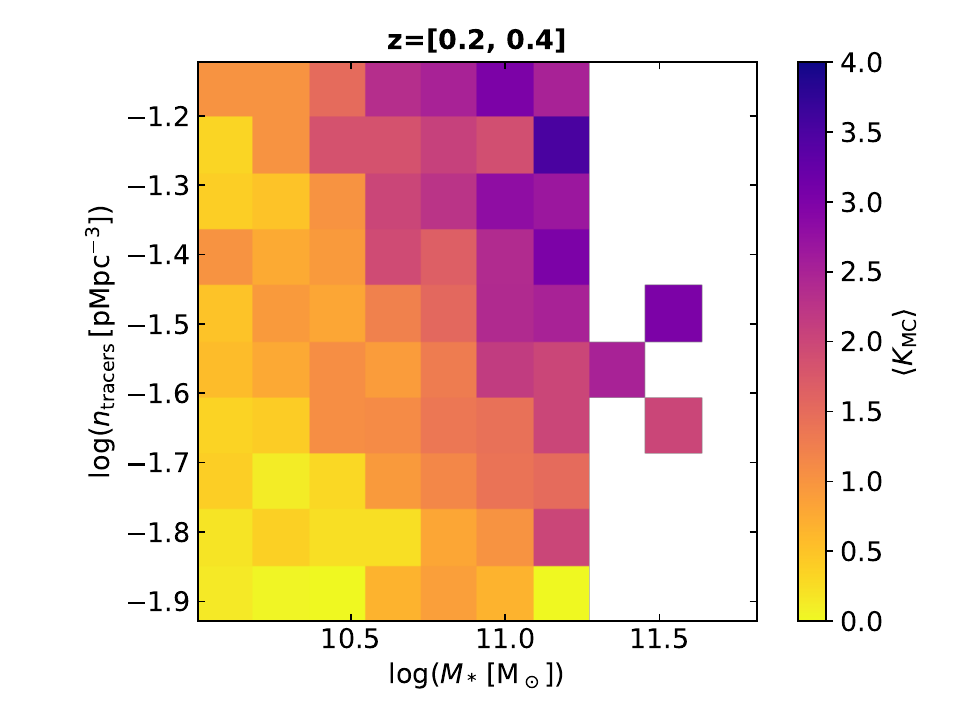}
    \includegraphics[width=0.45\textwidth]{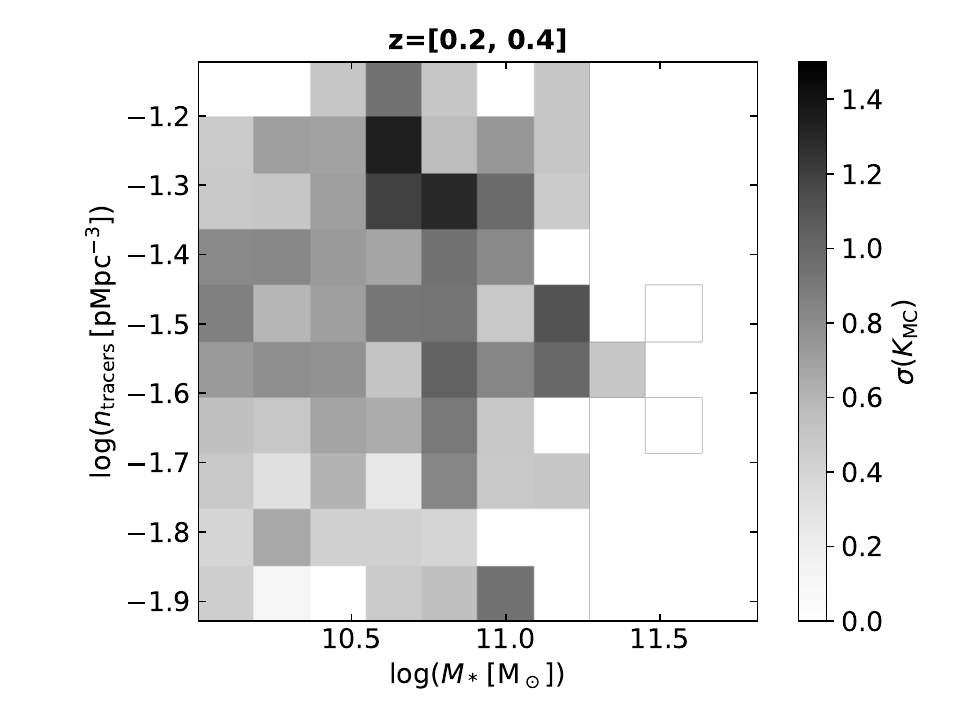}
    \includegraphics[width=0.45\textwidth]{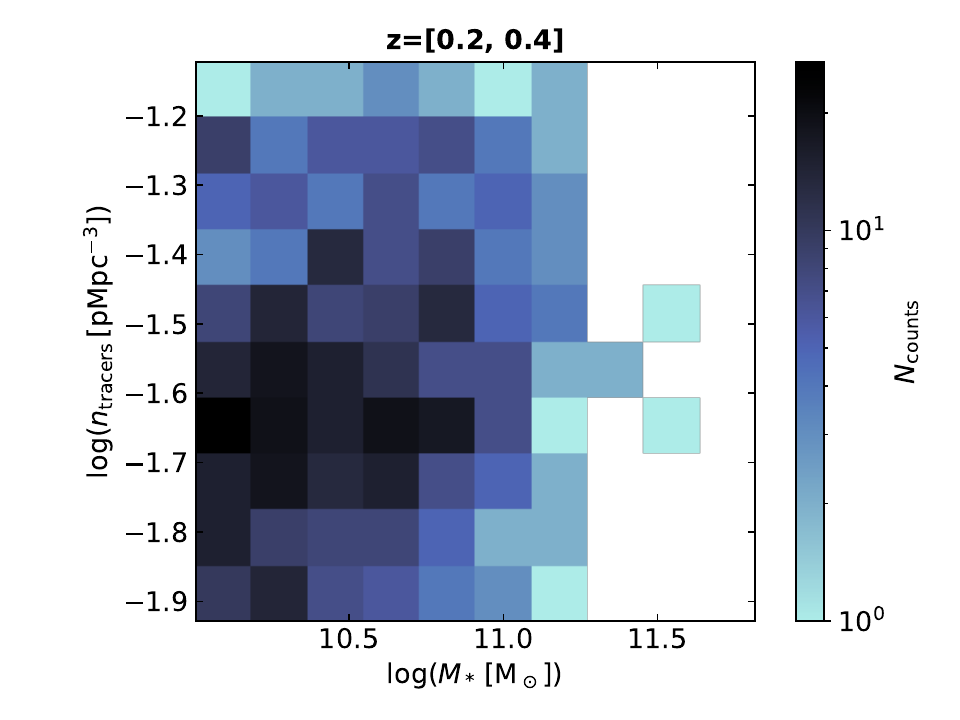}
   \caption{
   \textit{Top:} Connectivity of JPAS+DESI target galaxies in the mass-density plane for the low redshift subsample. Pixel colours represent the mean connectivity, $\langle K_\mathrm{MC} \rangle$, computed within each small bin of the mass–density space. \textit{Middle:} Standard deviation of the connectivity values within the same bins. \textit{Bottom:} Number of galaxies per bin.
   }
    \label{Fig:K_mass_density_plane}
    \end{figure}

Figure~\ref{Fig:Kcomparisons_1row} shows the resulting $K_\mathrm{MC}$ measurements as a function of galaxy stellar mass for different redshift bins (panels). The redshift bins were constructed to contain approximately the same number of galaxies. Results from the observed JPAS+DESI sample are shown as thick black lines, while the corresponding measurements for the mock galaxies, introduced in Sect.~\ref{Sect:mocks}, are shown in magenta. The error bars correspond to the $1\sigma$ dispersion around the mean value of the galaxies within the mass bin. We find a good agreement between the observed and mock trends, particularly at high redshift, reinforcing the reliability of the mocks as faithful representations of the observational data. Already at this stage, we see a clear trend of increasing connectivity with stellar mass, a point we will return to in more detail later on.\\

To assess the robustness of our probabilistic filament reconstruction and connectivity measurements, we compare the $K_\mathrm{MC}$ values with reference connectivities obtained for the same mock galaxies under ideal conditions in full 3D space. These reference reconstructions were performed on the \cite{Henriques2015_Lgal} lightcone using all available simulated galaxies, after applying a lower stellar mass cut of $M_\star > 10^{8.5}$~\Msun~ to remain well above the simulation’s resolution limit. We applied DisPerSE in Cartesian (physical) space within 3 pMpc radius spheres centred on each target galaxy, adopting a persistence threshold of $3\sigma$ for consistency with the Monte Carlo realisations and to ensure an equivalent detection level relative to the underlying Poisson noise. The resulting connectivities, $K_\mathrm{3D,true}$, (shown in blue in Fig.~\ref{Fig:Kcomparisons_1row}), serve as our true reference values, since they are unaffected by selection biases, projection effects, or redshift uncertainties.

The $K_\mathrm{MC}$ measurements show excellent agreement with the 3D reference values within the $1\sigma$ range at low redshift, where the density of tracers is the highest and miniJPAS photometric redshift uncertainties are the smallest. At higher redshifts, starting from the third panel, the MC reconstructions tend to yield slightly lower connectivities than the 3D reference. This bias likely arises because of the sparser sampling of tracers and the larger redshift errors, which makes it harder to capture the full filamentary signal. Nevertheless, the general trend of increasing connectivity with stellar mass is well recovered across all redshift bins. For a more detailed comparison, where galaxies are subdivided into different tracer density bins, we refer the reader to Fig.~\ref{Fig:APPKcomparisons} in Appendix~\ref{APP:Additional_material}.

Given these comparisons to reference measurements, in the remaining sections of this paper we restrict our analysis to the low redshift galaxies in our catalogue, $z = [0.2, 0.4]$, where the filament reconstructions are the most robust and capture the true connectivity of galaxies.

\subsection{\label{SubSect:Kmeasurements}Galaxy connectivity in the fundamental mass-density plane}

We now examine the dependence of galaxy connectivity on two fundamental parameters: stellar mass and tracer density. The top panel of Fig.~\ref{Fig:K_mass_density_plane} displays the connectivity of JPAS+DESI target galaxies across the mass–density plane in the restricted redshift range $z=[0.2, 0.4]$ (541 galaxies). The colour scale indicates the mean connectivity, $\langle K_\mathrm{MC} \rangle$, computed in small bins of this two-dimensional space, making it possible to visualise in detail how connectivity varies jointly with mass and density. 
The middle panel of this figure shows the corresponding $1 \sigma$ standard deviations, $\sigma(K_\mathrm{MC})$, which are generally small and show no clear correlation with the mean values above, supporting the significance of the trends discussed in the following. For reference, we show, in the bottom panels, the associated number counts per bin.

In the top panel of Fig.~\ref{Fig:K_mass_density_plane} we detect a clear correlation of connectivity with stellar mass, as the highest $K_\mathrm{MC}$ values (of around three connected filaments) are measured for the most massive galaxies. This trend is expected given the theoretical predictions of \cite{Codis2018}. It also agrees with the measurements from the hydro-dynamical simulations by \cite{GalarragaEspinosa2023}, although at higher redshift ($z = 2$) and using filaments traced directly in the dark matter density. For interpretations on why connectivity increases with mass, we refer the reader to those studies and the references therein.

\begin{figure}
    \centering
    \includegraphics[width=0.45\textwidth]{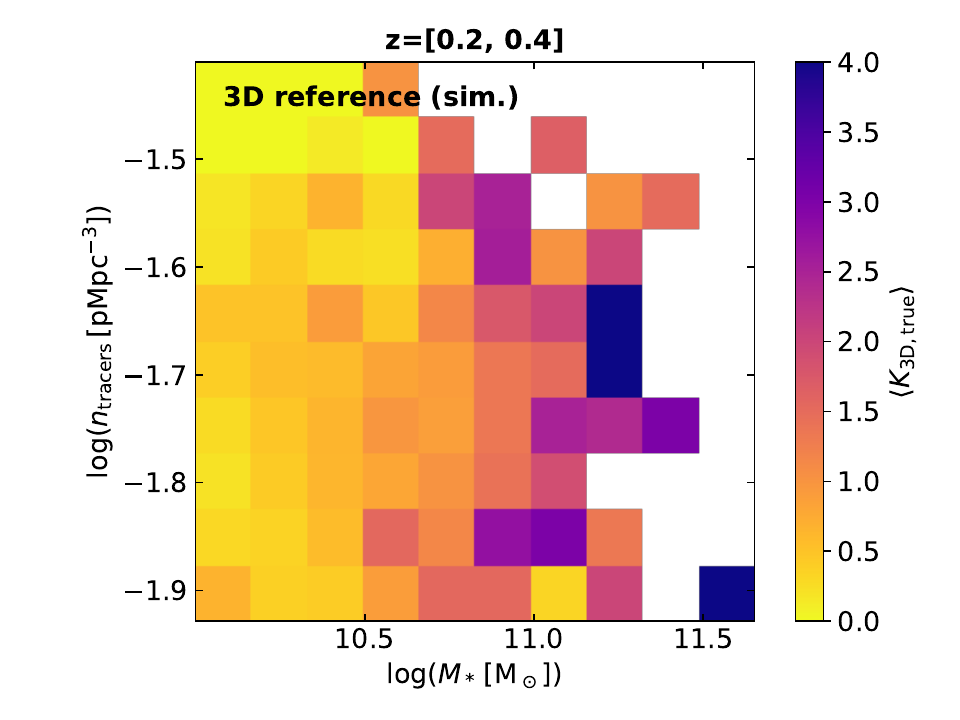}
    \caption{Same as Fig.~\ref{Fig:K_mass_density_plane} (top panel) but for our 3D reference connectivity measurements using the full lightcone galaxy distribution.
    }
    \label{Fig:K_plane_3Dref}
\end{figure}

We also observe a trend with tracer density, where, at fixed mass, connectivity increases as \ntracer~ rises. However, by comparing with our 3D reference measurements (Fig.~\ref{Fig:K_plane_3Dref}), we find that this trend originates from an observational bias, i.e. filaments in observations are more easily detected in regions with higher galaxy sampling. Indeed, the results of Fig.~\ref{Fig:K_plane_3Dref} using $K_\mathrm{3D,true}$ for galaxies in the same redshift range (0.2, 0.4]) show the opposite behaviour from observations, as connectivity decreases as tracer density increases. This is in agreement with previous theoretical studies based on full snapshots of hydro-dynamical simulations. For example, the analysis of the TNG50 simulation at $z=2$ of \cite{GalarragaEspinosa2023} reported that galaxies in denser and more crowded environments tend to show fewer filament connections than galaxies of the same mass in more relaxed, lower-density regions, due to the effect of tidal fields and interactions that tend to disconnect galaxies from their local web.
We therefore caution that the observed increase of connectivity with tracer density should not be over-interpreted, it mainly reflects the greater ease of identifying elongated structures when more tracers are available.

In light of these results, in the following we further subdivide our galaxy sample into narrow stellar mass and density bins, thereby controlling for the variations of connectivity with these parameters.

\subsection{\label{SubSect:SFR}Impact on galaxy star-formation rate}

We now investigate the potential influence of galaxy connectivity on star formation activity. To focus on the role of filaments alone, we exclude galaxies residing in groups and clusters in order to avoid mixing with environmental effects typical of dense group and cluster environments. We select these galaxies are those with a probability higher than 0.7 of belonging to one of the groups identified by AMICO in the miniJPAS field \citep{Maturi2023_jpasclusters}. Within $0.2 < z < 0.4$, there are 47 AMICO groups with estimated masses in the range $M_{200c} = 10^{12.8} - 10^{13.5}$~\Msun~. This step excludes 78 galaxies.

\begin{figure}
    \centering
    \includegraphics[width=1\linewidth]{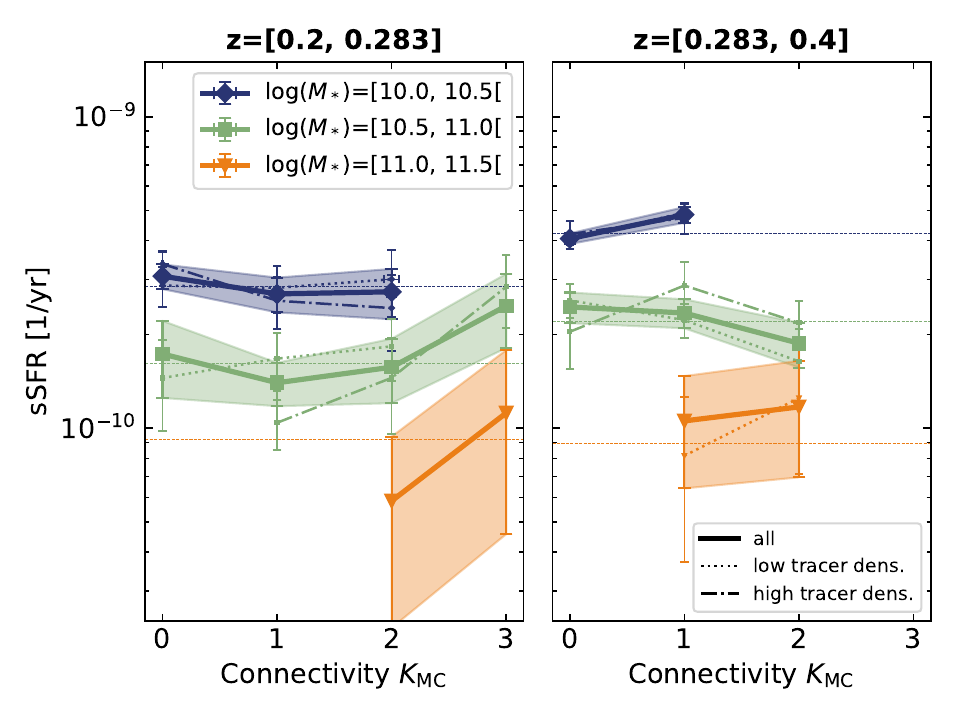}
    \caption{Specific star formation rate (sSFR) as a function of galaxy connectivity for the observed JPAS+DESI galaxies, shown across different redshift and stellar mass bins. Different line styles show galaxies in different tracer density bins. Galaxies in groups and clusters are excluded. Each point represents the mean sSFR at a given $K_\mathrm{MC}$ value; points based on fewer than three galaxies are not shown. Thin dashed horizontal lines indicate the average sSFR of the overall population in each mass bin. Error bars correspond to the standard errors of the mean, i.e. the standard deviation normalised by the square root of the number of data points in that bin.}
    \label{Fig:sSFRvsK}
\end{figure}

Figure~\ref{Fig:sSFRvsK} shows the relation between specific star formation rate (sSFR, defined as the ratio of a galaxy's star formation rate to its stellar mass) and connectivity, $K_\mathrm{MC}$. Galaxies are divided into two redshift bins (columns), and further split into narrow stellar mass bins (colours) and density bins (line styles). The number of galaxies in each redshift and mass bin is listed in Table~\ref{Table:Ngal_for_sSFR}. Thin dashed horizontal lines mark the average sSFR of galaxies in each mass bin. As expected, lower-mass galaxies are, on average, more star-forming than their massive counterparts.

\begin{table}[]
    \caption{Number of galaxies in the different redshift and stellar mass bins of Fig.~\ref{Fig:sSFRvsK}.}
    \label{Table:Ngal_for_sSFR}
    \centering
    \begin{tabular}{ c | c c c c c }
    \hline \hline
    $ \log(M_\star\,\, [\mathrm{M}_\odot])$ & z=[0.2, 0.283[ & z=[0.283, 0.4[ \\ \hline
    $[10.0, 10.5[$ & 86 & 151 \\
    $[10.5, 11[$ & 74 & 113 \\ 
    $[11, 11.5[$ & 11 & 27 \\
    $[11.5, 11.7[$ & 0 & 1 \\  \hline
    \end{tabular}
\end{table}

Focusing on the sSFR–$K_\mathrm{MC}$ relation, we find a mild trend of increasing sSFR with $K_\mathrm{MC}$, particularly for the lowest redshift, medium-mass galaxies. This may indicate enhanced star formation in systems with more filament connections. For other bins, however, the relation is flatter or even slightly negative. When galaxies are further divided by density, the results remain broadly consistent with the general trends. Given the limited sample size, at this stage it is difficult to establish conclusive results on the role of local filaments on galaxy star-formation activity. This will come from a future study on the full J-PAS footprint, using the same methods as those introduced in this work.

Before discussing our findings in the next section, we note that we do not attempt to reproduce the observed $K$–sSFR relation using the mock dataset. This is because the \cite{Henriques2015_Lgal} version of the L-Galaxies semi-analytic model includes only weak environmental prescriptions and does not account for filamentary gas accretion or coherent large-scale gas flows. Instead, we refer to past work that investigated this question using hydrodynamical simulations which follow the coevolution of dark matter, gas, and galaxies in a self-consistent way. In \cite{GalarragaEspinosa2023}, we reported the existence of a positive $K$–sSFR correlation at $z=2$, and ongoing work is extending this analysis to lower redshifts for a direct theoretical comparison with the present study.

\section{\label{Sect:discussion}Discussion}

This section discusses important points needed to interpret our findings. We focus first on the scales of the detected filaments within the multi-scale cosmic web, and then on the impact of filament connectivity on galaxy star formation.

\subsection{\label{SubSect:type_filaments}What type of filaments are we detecting?}

\begin{figure*}
    \centering
    \includegraphics[width=1\linewidth]{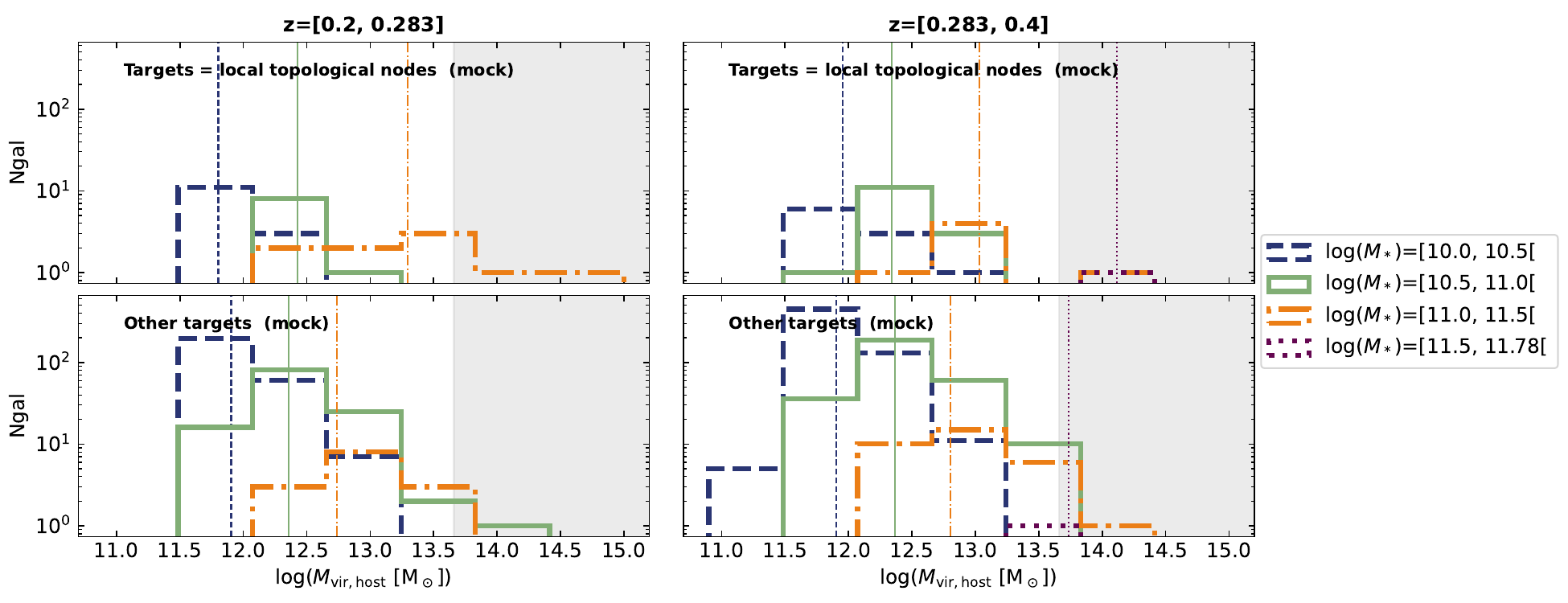}
    \caption{Halo mass distributions of the mock target galaxies, in the same redshift range as in the observations (different columns). The top panels focus on target galaxies that have been identified as local topological nodes of their field (see Sect.~\ref{SubSect:type_filaments}), while bottom panels show the distributions for the remaining targets, i.e. those who are not local nodes. The different colours represent different bins of stellar mass. Vertical thin lines show the median values of the distributions. The gray shaded area show the mass range of typical large-scale cosmic nodes (e.g. hosting galaxy clusters), taken from of \cite{GalarragaEspinosa2024_MTNG}.}
    \label{fig:M200_hosthalo}
\end{figure*}

The cosmic web is inherently multiscale, so interpreting our findings requires an understanding of the characteristic scales of the filaments we detect. One way to approach this is by examining the virial masses of the haloes hosting our target galaxies and comparing them to those of large-scale cosmic nodes, i.e. haloes connected to prominent filaments in the large-scale web.

Using the mock catalogue, we analyse in Fig.~\ref{fig:M200_hosthalo} the distribution of virial masses ($M_\mathrm{vir, host}$) of the haloes hosting the target galaxies, split into stellar mass bins.
In this figure, we distinguish between targets that are likely local topological nodes of their field (top panels) and those that are not (bottom panels). The probability of being a local node is derived from the Monte Carlo filament reconstructions: in each realisation, a target is identified as a local node if it corresponds to a maximum density critical point within 700 pkpc of its position.\footnote{Using 500 pkpc yields consistent results}. Galaxies considered as local nodes are those with  $>50\%$ probability.

We find that the host haloes of our targets typically have virial masses below the characteristic range of large-scale cosmic nodes, as estimated by \cite{GalarragaEspinosa2024_MTNG} (grey shaded regions). Only at the highest stellar masses ($M_\star \gtrsim 10^{11}$~\Msun~) do the host halo masses approach the cosmic node regime. This is consistent with expectations, as these massive galaxies are likely the central, brightest objects in cluster-sized haloes. \\

A natural next question is whether our target galaxies are themselves embedded within particular structures of the large-scale cosmic web. For instance, if a galaxy lies within a large-scale filament, the structures detected here could correspond to segments of that larger feature. Conversely, if a galaxy resides in a wall or void, it may instead be a local node of its own smaller-scale network of `primordial' filaments \citep[e.g.][]{Alpaslan2014_tendrils, Borzyszkowski2017_ZOMG1, AragonCalvo2019_CWdisconnection}.

The upper panels of Fig.~\ref{fig:M200_hosthalo}, which focus on targets identified as local nodes, show that galaxies with stellar masses between $10^{10}$ and $10^{11}$~\Msun~ are unlikely to coincide with large-scale cosmic nodes. Instead, they appear to be nodes of their own, smaller-scale filamentary network. The filaments connected to these galaxies therefore most likely trace finer structures, on scales smaller than the typical $\sim1$~pMpc width of large-scale cosmic filaments \citep{Cautun2014, GalarragaEspinosa2024_MTNG, Bahe2025_filaments}, and arising from the local topology of the density field rather than the global web.
Galaxies that are not local nodes, on the other hand, are most plausibly embedded within or adjacent to a larger-scale filament, and they could be tracing its skeleton without being a point of convergence themselves. This might not, however, preclude them from accreting material from those filaments \citep{Borzyszkowski2017_ZOMG1}.

Ultimately, answering this question definitively requires a full reconstruction of the large-scale cosmic web to characterise the broader environments of the targets. At present, this is not feasible given the limited field of view of the miniJPAS survey. Future analyses using the full J-PAS dataset will overcome this limitation, allowing for a comprehensive mapping of the large-scale structure and a more complete understanding of the multiscale web surrounding these galaxies.

\subsection{On the sSFR--$K$ relation}

In Sect.~\ref{SubSect:SFR}, we found tentative evidence that the sSFR of galaxies increases with higher connectivity. This trend is consistent with the scenario in which cosmic filaments act as efficient conduits for gas accretion onto galaxies \citep[anisotropic gas accretion, as predicted by hydrodynamical simulations, e.g.][although at higher redshift]{Keres2005, Dekel2009_coldstreams, Ramsoy2021}. In such a picture, a higher number of filament connections may provide additional pathways for the inflow of cold gas, thereby sustaining or enhancing star formation activity. This aligns with the predictions of the cosmic web detachment (CWD) model \citep{AragonCalvo2019_CWdisconnection}, in which galaxies that remain connected to the web continue to accrete gas and form stars, while those that detach become quenched.

As discussed in Sect.~\ref{SubSect:SFR}, the current analysis is strongly limited by sample size. The forthcoming extension of this study using the full J-PAS dataset will significantly improve the statistics and allow a more detailed and conclusive exploration of the sSFR–connectivity relation.

\section{\label{Sect:Conclusions}Conclusions}

In this work, we present the first systematic detection of the local cosmic web around galaxies spanning the stellar mass range $10^{10} -10^{11.7}$ \Msun~ at redshifts 0.2 to 0.8. By combining galaxies from the spectro-photometric miniJPAS survey with DESI spectroscopic redshifts (Sect.~\ref{Sect:jpas+desi_catalogue}), we created a unique dataset with dense sampling of tracers and attempted a detection of filaments in the 3 pMpc environments around individual target galaxies. Filament detection was made possible by adopting a probabilistic Monte Carlo reconstruction method (Fig.~\ref{Fig:OBS_MCvisu}, Sect.~\ref{Sect:Method}) that mitigates the effects of photo-z uncertainties, enabling reliable measurements of galaxy connectivity. Throughout this work, we used galaxy connectivity, $K$, as the main metric to quantify the environments of the target galaxies. Connectivity is defined as the number of filaments intersecting a mass-dependent radius, chosen to match the typical CGM scales of each individual target galaxy.

We paid particular attention to assessing potential systematics in our method to ensure the robustness of our results. A crucial part of this effort was the construction of a mock galaxy catalogue that replicates the selection function applied to the observations (Sect.~\ref{Sect:mocks}), allowing for comparison with reference 3D results for the same galaxies. 
A summary of our main results and interpretations, presented in Sect.~\ref{Sect:Results} and Sect.~\ref{Sect:discussion}, is provided below:

\begin{itemize}
    \item The connectivity measurements obtained with our probabilistic Monte Carlo filament detection method, $K_\mathrm{MC}$, show good agreement with ideal 3D measurements (Fig.~\ref{Fig:Kcomparisons_1row}), demonstrating that, in the low redshift regime ($z<0.4$), our filament reconstructions are robust against redshift uncertainties and selection effects in the observed galaxy catalogue.

    \item We analysed the scales of the recovered filaments (Sect.~\ref{SubSect:type_filaments}) and found that, except for the most massive galaxies (\Mstar~ $> 10^{11}$~\Msun~), the detected filaments are unlikely to trace large-scale cosmic structures. Instead, they are more likely associated with local structures shaped by the topology of the surrounding density field, forming a local web rather than a global one.
    
    \item By separating galaxies into narrow stellar mass bins and excluding systems in groups and clusters, we investigated the impact of filament connectivity on galaxy sSFR. We found a tentative positive correlation in our low redshift bins (Fig.~\ref{Fig:sSFRvsK}), but this needs to be confirmed by follow-up studies with larger sample sizes. If confirmed, these trends would support a scenario in which increased connectivity provides more channels for the inflow of cold gas that fuels star formation.
    
\end{itemize}

Unlike traditional density-based environment metrics (e.g., local overdensity), galaxy connectivity opens the door to measuring the anisotropy of the underlying density field, thus allowing to probe anisotropic inflows of matter towards haloes. Galaxy connectivity offers a new perspective on galaxy evolution and complements typical metrics used to study environmental effects.\\

In conclusion, this work represents a significant step forward in detecting the small-scale cosmic web and understanding the connection between galaxies to their local anisotropic environment, as opposed to previous literature that has focused on connections to the large-scale web. In the near future, we plan to expand this analysis to the larger sky coverage of the ongoing J-PAS, which is crucial for enhancing the statistical power of our results. Additionally, to refine our interpretations, we intend to conduct theoretical studies using state-of-the-art hydrodynamical simulations at these low redshift ranges. These studies will focus on the gas content and potential biases of the filaments traced by galaxies, compared to those identified directly in the local dark-matter density around targets, particularly in the low-mass regime.

Current and upcoming spectroscopy surveys like the Prime Focus Spectrograph (PFS), Euclid, DESI, and 4MOST (with the 4HS and WAVES surveys) will provide the necessary data to not only explore the sSFR–$K$ relation but also, thanks to their enhanced statistical power, to extend these types of analyses across a broader parameter space, including morphology, angular momentum, metallicity, among others. These future datasets will be crucial in advancing our understanding of galaxy evolution within a cosmological context that considers the multiscale nature of the environments in which galaxies form and evolve.

\begin{acknowledgements}
    We thank the anonymous referee for providing useful comments and suggestions that improved the paper. DGE thanks Khee-Gan Lee, and Mohammadreza Ayromlou for helpful discussions, and Clotilde Laigle for the useful and very thorough comments on the draft leading to the final version of the paper. SB acknowledge support from the Spanish Ministerio de Ciencia e Innovación through project PID2021-124243NB-C21. LLS acknowledges support by the Deutsche Forschungsgemeinschaft (DFG, German Research Foundation) under Germany's Excellence Strategy – EXC 2121 'Quantum Universe' – 390833306. RGD acknowledges financial support from the Severo Ochoa grant CEX2021-001131-S, funded by MICIU/AEI (10.13039/501100011033), and is also grateful for support from project PID2022-141755NB-I00. ET was funded by the HTM (grant TK202), ETAg (grant PRG1006) and the EU Horizon Europe (EXCOSM, grant No. 101159513). SGL acknowledges the financial support from the MICIU with funding from the European Union NextGenerationEU and Generalitat Valenciana in the call Programa de Planes Complementarios de I+D+i (PRTR 2022) Project (VAL-JPAS), reference ASFAE/2022/025. VM thanks CNPq (Brazil) and FAPES (Brazil) for partial financial support. This work is part of the research Project PID2023-149420NB-I00 funded by MICIU/AEI/10.13039/501100011033 and by ERDF/EU. This work is also supported by the project of excellence PROMETEO CIPROM/2023/21 of the Conselleria de Educación, Cultura, Universidades y Empleo (Generalitat Valenciana).
    This work is based on observations made with the JST250 telescope and PathFinder camera for the miniJPAS project at the Observatorio Astrof\'{\i}sico de Javalambre (OAJ), in Teruel, owned, managed, and operated by the Centro de Estudios de F\'{\i}sica del  Cosmos de Arag\'on (CEFCA). We acknowledge the OAJ Data Processing and Archiving Unit (UPAD; Cristobal-Hornillos et al. 2012) for reducing and calibrating the OAJ data used in this work.
    Funding for the J-PAS Project has been provided by the Governments of Spain and Arag\'on through the Fondo de Inversiones de Teruel; the Aragonese Government through the Research Groups E96, E103, E16\_17R, E16\_20R, and E16\_23R; the Spanish Ministry of Science and Innovation (MCIN/AEI/10.13039/501100011033 y FEDER, Una manera de hacer Europa) with grants PID2021-124918NB-C41, PID2021-124918NB-C42, PID2021-124918NA-C43, and PID2021-124918NB-C44; the Spanish Ministry of Science, Innovation and Universities (MCIU/AEI/FEDER, UE) with grants PGC2018-097585-B-C21 and PGC2018-097585-B-C22; the Spanish Ministry of Economy and Competitiveness (MINECO) under AYA2015-66211-C2-1-P, AYA2015-66211-C2-2, and AYA2012-30789; and European FEDER funding (FCDD10-4E-867, FCDD13-4E-2685). The Brazilian agencies FINEP, FAPESP, FAPERJ and the National Observatory of Brazil have also contributed to this project. Additional funding was provided by the Tartu Observatory and by the J-PAS Chinese Astronomical Consortium.
    \textbf{Author contributions:} 
    \textbf{DGE}: conceptualization; formal analysis; investigation; methodology; data curation; software; validation; visualization; writing - original draft, review \& editing.
    \textbf{GK}: conceptualization; methodology; validation; writing - review \& editing.
    \textbf{SB}: conceptualization; methodology; validation.
    \textbf{LLS}: methodology; writing - review \& editing.
    \textbf{RGD}: resources; data curation; writing - review \& editing.
    \textbf{ET}: methodology; writing - review \& editing.
    \textbf{RA}: resources; writing - review \& editing.
    \textbf{SGL}: writing - review \& editing.
    \textbf{VM, JA, NB, SC, JC, DCH, RD, AE, AHC, CHM, CLS, AMF, CMO, MM, LS, KT, JV, HVR}: resources.
\end{acknowledgements}

\bibliography{main} 

%
%

\appendix

\section{\label{APP:apertureradius}Choice of aperture in connectivity measurements}

In this appendix, we test the robustness of our connectivity measurements to variations in the aperture radius. We focus on galaxies in the same redshift range as in Fig.~\ref{Fig:K_mass_density_plane}, which are the galaxies with reliable filament identifications based on our comparisons to 3D reference measurements. 

\begin{figure}
    \centering
    \includegraphics[width=0.5\linewidth]{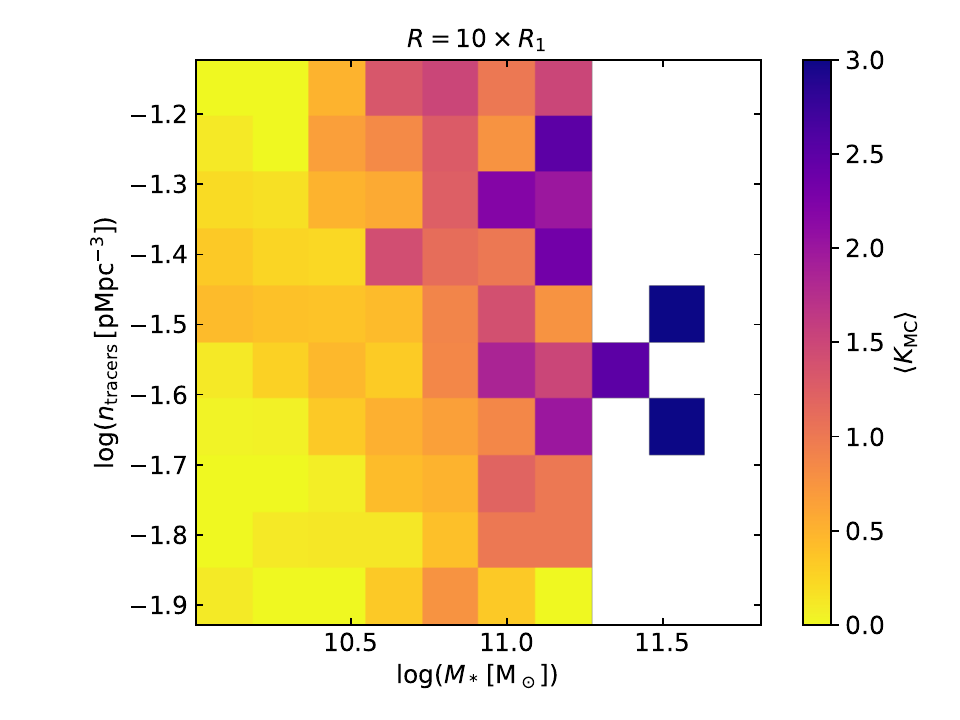}\includegraphics[width=0.5\linewidth]{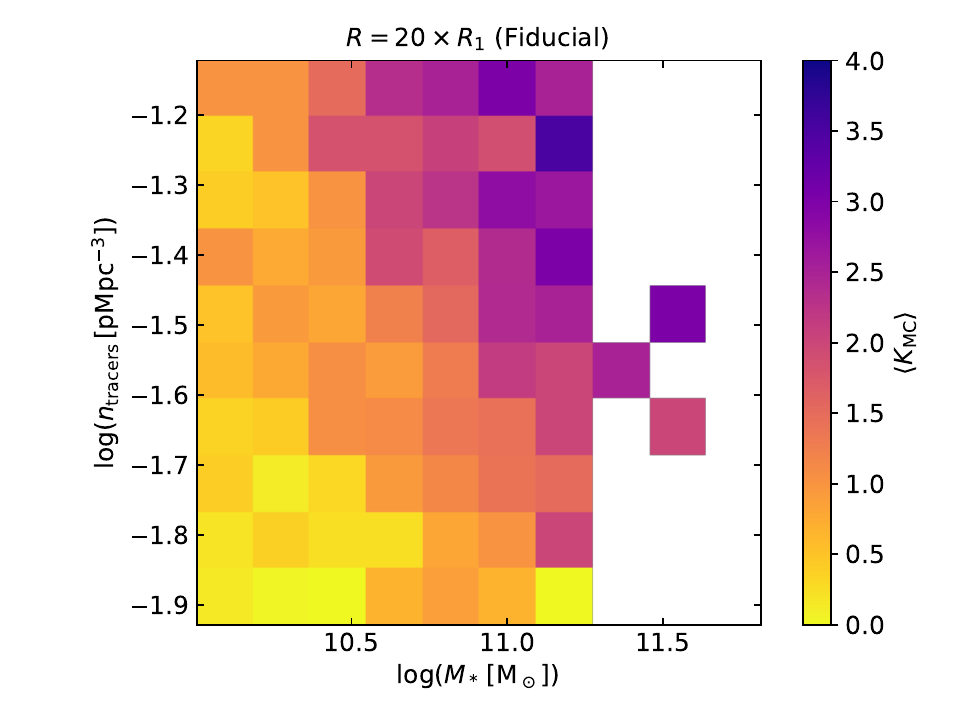}
    \includegraphics[width=0.5\linewidth]{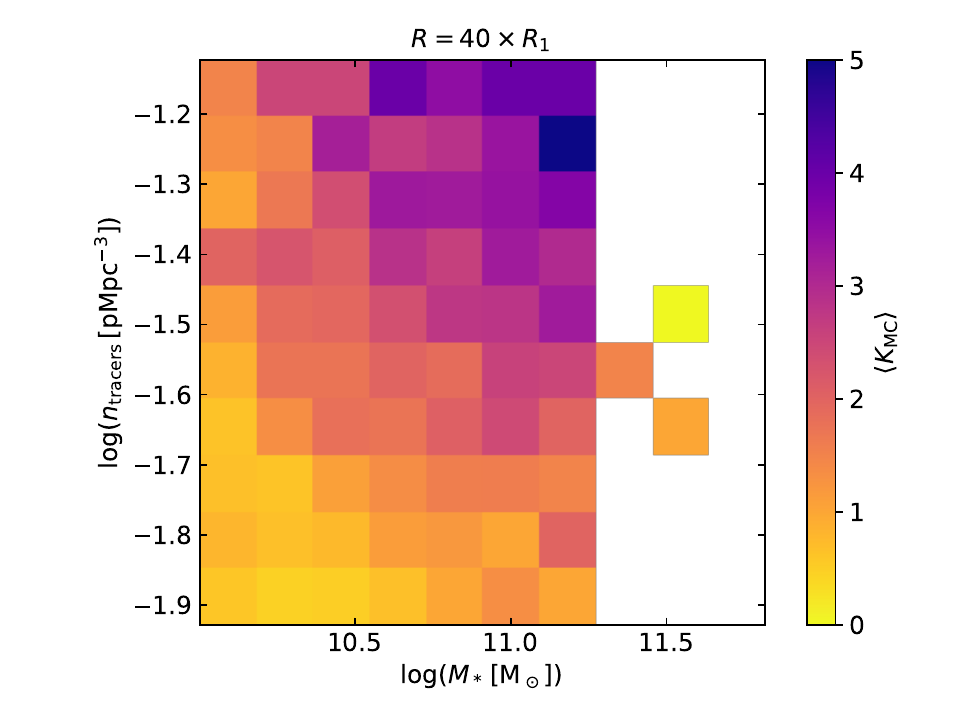}\includegraphics[width=0.5\linewidth]{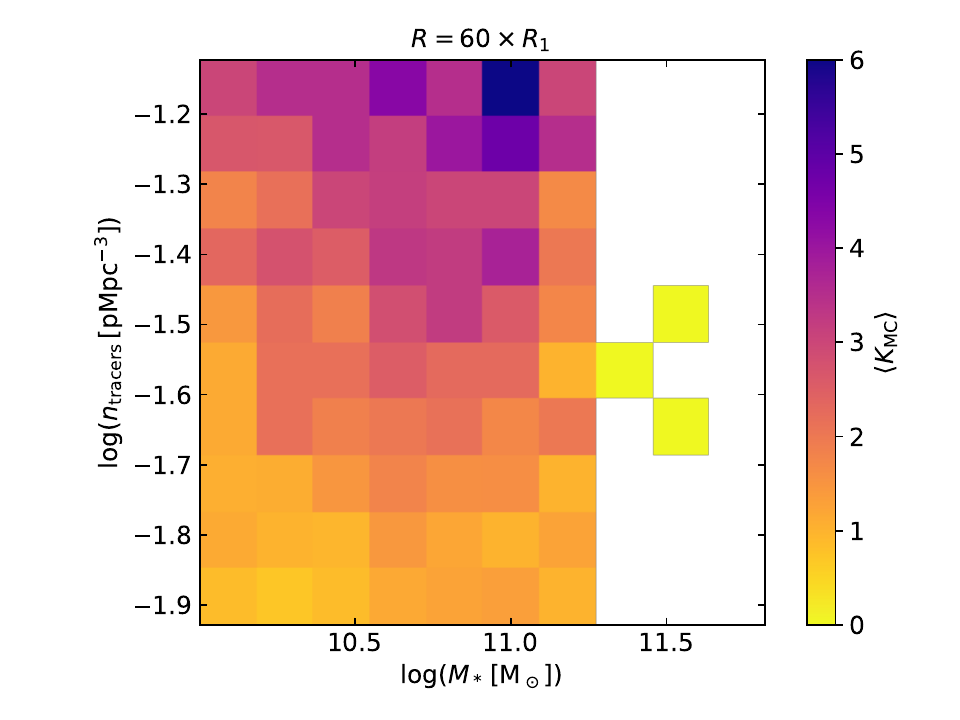}
\caption{Connectivity in the mass-density plane resulting from different choices of apertures. Here we vary the factor associated to the $R_1$ radius (see Sect~\ref{SubSect:MeasuringK_method} for details on this radius).}
    \label{Fig:Kaperture_tests}
\end{figure}

Figure~\ref{Fig:Kaperture_tests} shows the results obtained when varying the $R_1$ factor radius. We explore factors of $10, 20, 40$ and $60 \times R_1$, respectively, from left to right panels.
For factors of 20 and above, we find that the connectivity estimates and their trends with mass and density remain qualitatively consistent among them. Some differences appear for the most massive galaxies, where very large apertures (e.g. the rightmost panel) exceed the 3 pMpc radial limit, leading to artificially low values of $K_\mathrm{MC}=0$. At the other extreme, for the smallest aperture of $10 \times R_1$, the trends with mass and density are still recovered, but the number of galaxies with connectivity values of zero is significantly larger. This is expected, as such a small aperture makes any connectivity signal difficult to capture.\\

\begin{figure}
    \centering
    \includegraphics[width=0.5\linewidth]{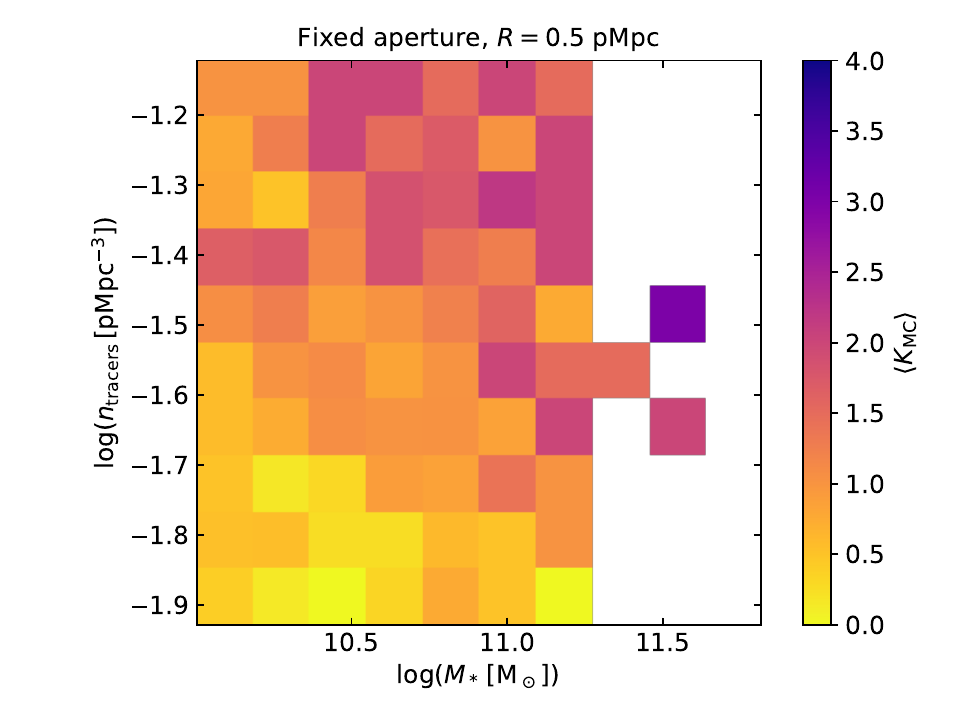}\includegraphics[width=0.5\linewidth]{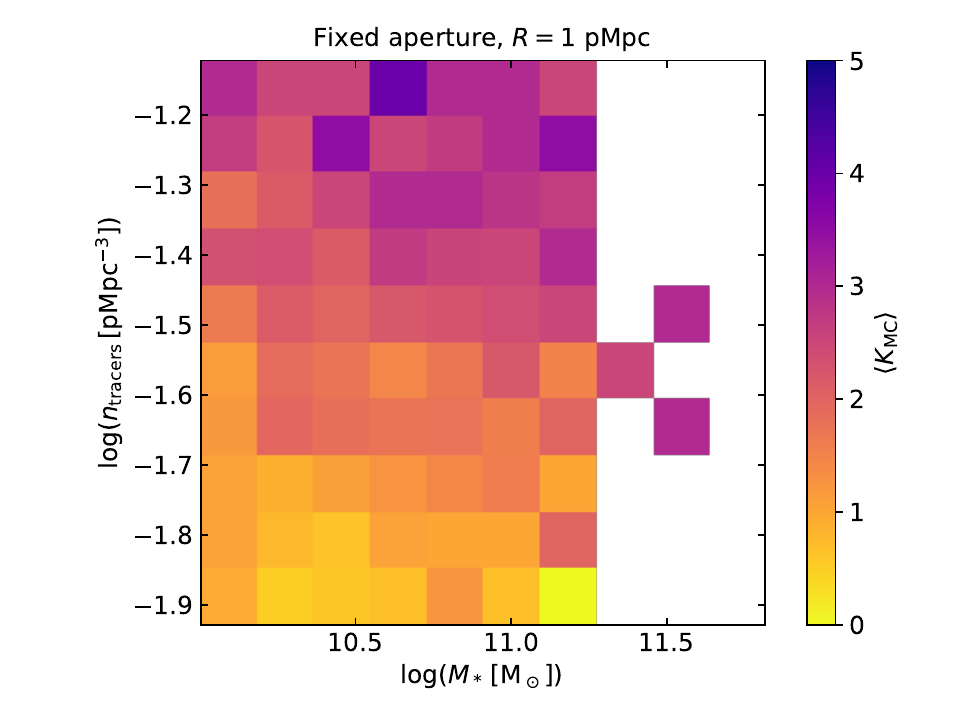}
    \includegraphics[width=0.5\linewidth]{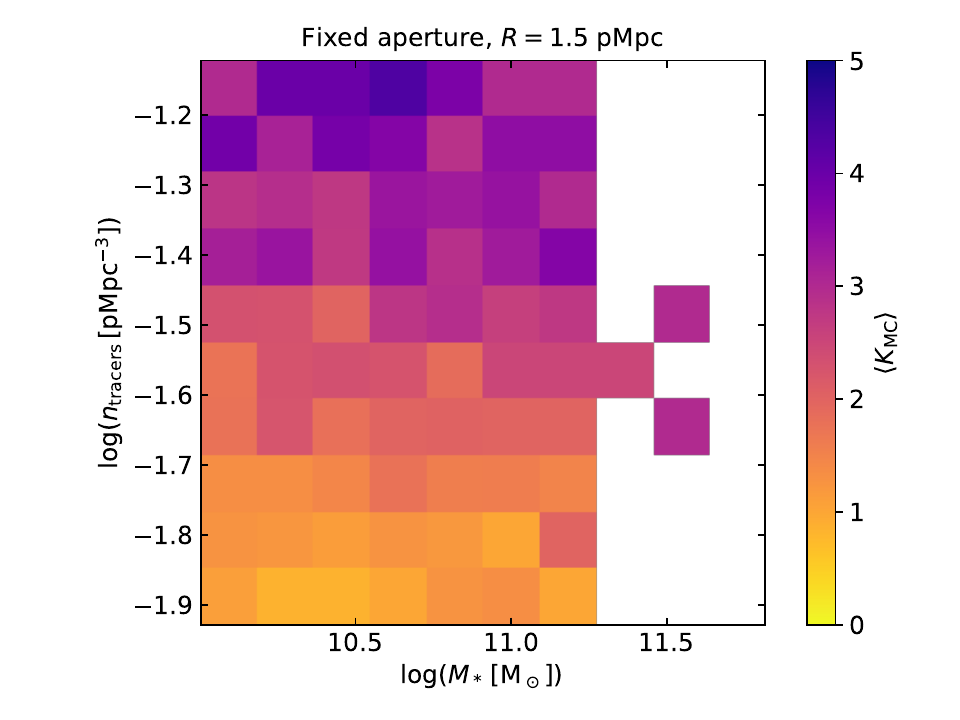}
\caption{Same as Fig.~\ref{Fig:Kaperture_tests} but when connectivity is measured using fixed aperture radii, constant for all galaxy masses.
    }
    \label{Fig:Kaperture_tests_FIXED}
\end{figure}

We repeat the analysis using fixed aperture radii of 0.5, 1.0, and 1.5 pMpc, keeping the aperture size constant across the entire stellar mass range. Although this approach is not physically motivated—since a galaxy’s sphere of influence likely scales with its mass—it provides a useful consistency test to verify that the observed $K$–$M_\star$ relation is not an artefact of the mass-dependent apertures used in our fiducial analysis. The resulting connectivity distributions, for the same galaxy sample as in Fig.~\ref{Fig:K_mass_density_plane}, are shown in Fig.~\ref{Fig:Kaperture_tests_FIXED}.

We repeat the analysis using fixed aperture radii of 0.5, 1.0, and 1.5 pMpc. In these cases, the aperture radius is constant across the entire galaxy mass range. While this approach is hard to physically motivate, since a galaxy's scale of influence is most likely dependent on its mass, it serves as a useful test to confirm that the observed $K$–$M_\star$ trend is not simply driven by the use of mass-dependent apertures. The corresponding connectivity distributions for the same galaxies as in Fig.~\ref{Fig:K_mass_density_plane} are shown in Fig.~\ref{Fig:Kaperture_tests_FIXED}.

The trend of increasing connectivity with stellar mass is still present for the 0.5 and 1.0 pMpc apertures (top panels), although slightly weaker than in the fiducial case. However, for the largest aperture of 1.5 pMpc (bottom), the trend with mass nearly disappears. This behaviour is expected when the aperture becomes too large, since it effectively smooths out any signal from the low mass target galaxies, as the measurement is done well beyond the physical environment directly associated with the target galaxy. As a result, the measured connectivity instead reflects variations in the overall tracer density rather than intrinsic differences between galaxies.

These results support the robustness of the $K$–$M_\star$ relation to variations in the aperture definition and confirm that the observed trends in this paper are not artifacts of our chosen parameterisation.

\section{\label{APP:Additional_material}Additional material on the comparison with 3D connectivity measurements}

\begin{figure*}
    \centering
    \includegraphics[width=1\linewidth]{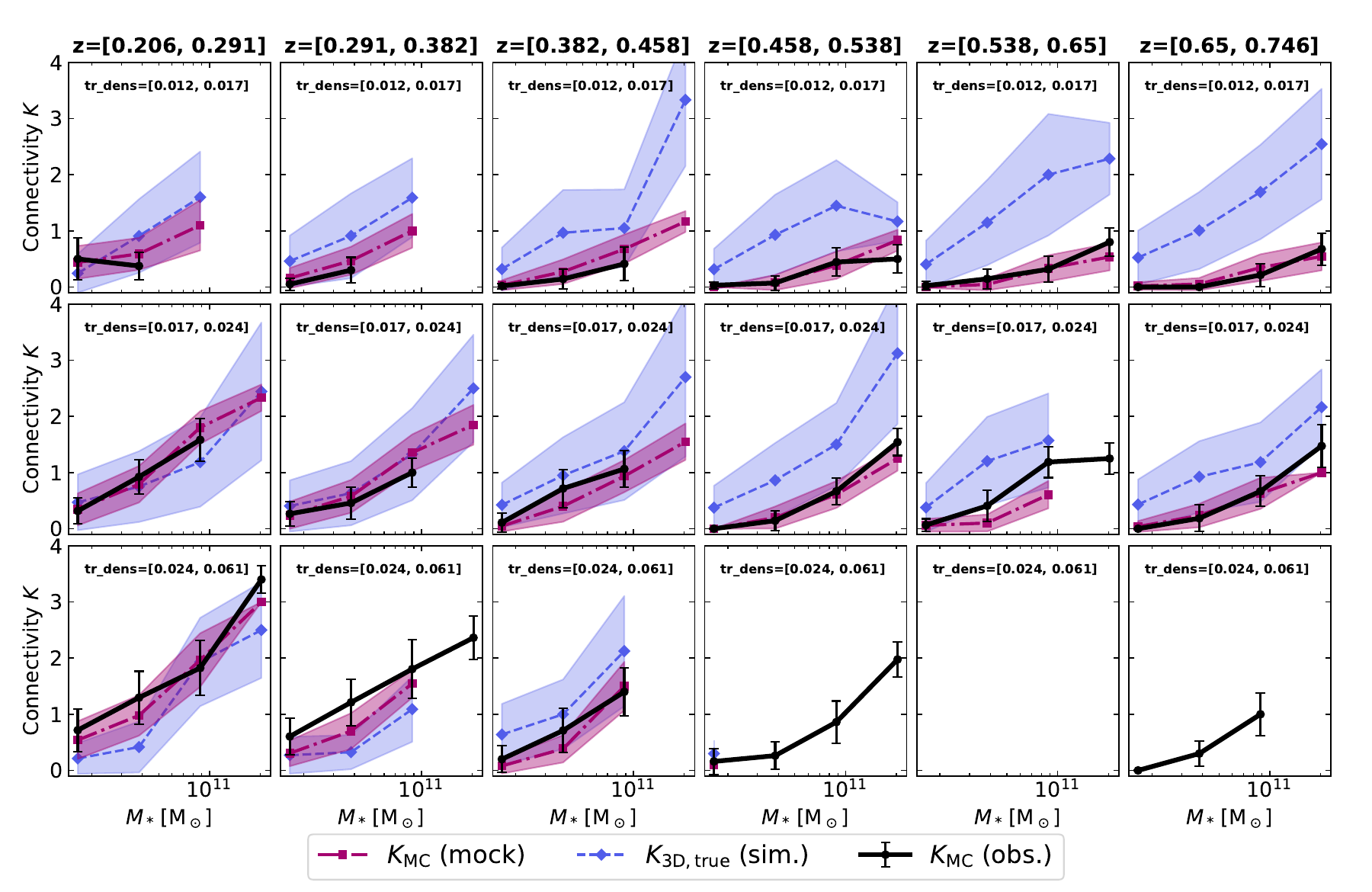}
    \caption{Same as Fig.~\ref{Fig:Kcomparisons_1row} but with galaxies further split in bins of increasing tracer density (different rows). Density bin edges were chosen to contain approximately the same number of observed galaxies. Averages based on fewer than four galaxies are not shown.
    }
    \label{Fig:APPKcomparisons}
\end{figure*}

We present in Fig.~\ref{Fig:APPKcomparisons} the results of our comparison to reference measurements where galaxies are further split in density bins. These results are consistent with the general trends reported in Fig.~\ref{Fig:Kcomparisons_1row}. 

\end{document}